\documentclass{mn2e}
\usepackage{epsfig}
\usepackage{amsmath}
\usepackage{amssymb}
\usepackage{hyperref}
\usepackage{verbatim}
\usepackage{natbib}
\usepackage{color}
\usepackage{multicol}
\begin{document}

\title[Directional Gravitational Wave Searches with PTAs]{Versatile Directional Searches for Gravitational Waves with Pulsar Timing Arrays}
\author[D. R. Madison et al.]
{D. R. Madison$^{1,2}$,
X.-J. Zhu$^{3,4}$,
G. Hobbs$^4$,
W. Coles$^5$,  
R. M. Shannon$^{4,6}$,
\newauthor
J. Wang$^7$,   
C. Tiburzi$^{8,9,10,11}$,
R. N. Manchester$^4$,
M. Bailes$^{12}$,
N. D. R. Bhat$^{6}$,
\newauthor
S. Burke-Spolaor$^{13}$,
S. Dai$^{14,4}$,
J. Dempsey$^{15}$,
M. Keith$^{16}$,
M. Kerr$^4$,
P. Lasky$^{17,18}$,
\newauthor
Y. Levin$^{17}$,
S. Os\l{}owski$^{11,10}$,
V. Ravi$^{19}$,
D. Reardon$^{17,4}$,
P. Rosado$^{12}$,
R. Spiewak$^{20}$,
\newauthor
W. van Straten$^{12}$,
L. Toomey$^4$,
L. Wen$^3$,
\& X. You$^{21}$.
\\ 
$^1$ Department of Astronomy, Cornell University, Ithaca, NY 14853, USA; drm252@cornell.edu\\
$^2$ National Radio Astronomy Observatory, 520 Edgemont Rd., Charlottesville, VA 22903, USA\\
$^3$ School of Physics, University of Western Australia, 35 Stirling Hwy, Crawley WA 6009, Australia\\
$^4$ CSIRO Astronomy and Space Science, Australia Telescope National Facility, PO Box 76, Epping NSW 1710, Australia\\
$^5$ Electrical and Computer Engineering, University of California at San Diego, La Jolla, California, USA\\
$^6$ International Centre for Radio Astronomy Research, Curtin University, Bentley, WA 6102, Australia\\
$^7$ Xinjiang Astronomical Observatory, CAS, 150 Science 1-Street, Urumqi, Xinjiang 830011, China\\	
$^8$ INAF - Osservatorio Astronomico di Cagliari, Via della Scienza, 09047 Selargius (CA), Italy\\
$^9$ Dipartimento di Fisica, Universit\`{a} di Cagliari, Cittadella Universitaria 09042 Monserrato (CA), Italy\\
$^{10}$ Max-Planck-Institut f\"{u}r Radioastronomie, Auf dem H\"{u}gel 69, 53121 Bonn, Germany\\
$^{11}$ Fakult\"{a}t f\"{u}r Physik, Universit\"{a}t Bielefeld, Postfach 100131, 33501 Bielefeld, Germany\\
$^{12}$ Centre for Astrophysics and Supercomputing, Swinburne University of Technology, P.O. Box 218, Hawthorn, VIC 3122, Australia\\
$^{13}$ National Radio Astronomy Observatory, 1003 Lopezville Rd., Socorro, NM 87801, USA\\
$^{14}$ Department of Astronomy, School of Physics, Peking University, Beijing 100871, China\\
$^{15}$ CSIRO Information Management \& Technology (IM\&T), PO Box 225, Dickson ACT 2602, Australia\\
$^{16}$ Jodrell Bank Centre for Astrophysics, School of Physics and Astronomy, The University of Manchester, Manchester M13 9PL, UK\\
$^{17}$ Monash Centre for Astrophysics (MoCA) School of Physics and Astronomy, Monash University, VIC 3800, Australia\\
$^{18}$ School of Physics, University of Melbourne, Parkville, VIC 3010, Australia\\
$^{19}$ California Institute of Technology, Pasadena, CA 91125, USA\\
$^{20}$ Department of Physics, University of Wisconsin-Milwaukee, P.O. Box 413, Milwaukee, WI 53201, USA\\
$^{21}$ School of Physical Science and Technology, Southwest University, Chongqing 400715, China
}
\maketitle

\begin{abstract}
By regularly monitoring the most stable millisecond pulsars over many years, pulsar timing arrays (PTAs) are positioned to detect and study correlations in the timing behaviour of those pulsars. Gravitational waves (GWs) from supermassive black hole binaries (SMBHBs) are an exciting potentially detectable source of such correlations. We describe a straight-forward technique by which a PTA can be ``phased-up" to form time series of the two polarisation modes of GWs coming from a particular direction of the sky. Our technique requires no assumptions regarding the time-domain behaviour of a GW signal. This method has already been used to place stringent bounds on GWs from individual SMBHBs in circular orbits. Here, we describe the methodology and demonstrate the versatility of the technique in searches for a wide variety of GW signals including bursts with unmodeled waveforms. Using the first six years of data from the Parkes Pulsar Timing Array, we conduct an all-sky search for a detectable excess of GW power from any direction. For the lines of sight to several nearby massive galaxy clusters, we carry out a more detailed search for GW bursts with memory, which are distinct signatures of SMBHB mergers.  In all cases, we find that the data are consistent with noise.
\end{abstract}

\begin{keywords}
pulsars--gravitational waves--data analysis
\end{keywords}

\section{Introduction}

Pulsar timing arrays (PTAs) provide a unique means for detecting gravitational waves (GWs) with frequencies between approximately $10^{-9}$ and $10^{-6}$ Hz. Sensitivity at such frequencies makes PTAs key for searching for and eventually studying GWs from supermassive black hole binaries (SMBHBs) with masses greater than $10^{7}~M_{\odot}$ \citep{sv10}.  PTAs are therefore indispensable for understanding galaxy evolution over cosmological timescales, for investigating the mechanisms by which the final parsec problem \citep{mm03} may be solved, and for probing strong gravitational fields.

The PTA concept was initially conceived decades ago \citep{d79,hd83,fb90}, and is now being realised by several international collaborations.  The European Pulsar Timing Array \citep[EPTA;][]{kc13}, the North American Nanohertz Observatory for Gravitational Waves \citep[NANOGrav;][]{m13_2}, and the Parkes Pulsar Timing Array \citep[PPTA;][]{mhb+13} collaborations use sensitive radio telescopes to observe the most rotationally stable millisecond pulsars (MSPs) known and measure the highest-precision pulse times of arrival (ToAs) possible on a regular basis.  Measured ToAs are compared to predictions of timing models that aim to account for all of the known physical effects that modulate the regularity with which pulses arrive at Earth-based observatories \citep{ehm06}.  After sufficiently long periods of time, the differences between measured ToAs and model predictions, the timing residuals, may begin to show structure indicative of errors in terrestrial time standards \citep{hcm+12}, incorrect solar-system ephemerides \citep{chm+10}, or the influence of GWs on the pulsar-Earth system. To analyse such effects, the various PTAs are also combining their data and expertise in the International Pulsar Timing Array \citep[IPTA;][]{haa+10,m13_1} which is poised to become the most sensitive tool for such investigations.

The GW signals potentially detectable by PTAs fall into two classes. First, individually resolvable sources such as SMBHBs in circular \citep{abb+14,zhw+14,bps+15} or eccentric orbits \citep{rws+14,hmg+15}, and burst sources, especially so-called ``bursts with memory" (BWMs) from the final merger of SMBHBs \citep{vl10,cj12,whc+15,abb+15} or potentially from exotic sources like phase transitions in the early Universe \citep{cbv+14}.  Second, an isotropic stochastic background (SB) of GWs created by the incoherent superposition of many unresolved SMBHBs scattered throughout the Universe \citep[][]{vlj+11,dfg+13,src+13,ltm+15,abb+15_2,srl+15}. An isotropic SB is an idealization and recent work has begun to place limits on anisotropic features of the background \citep{tmg+15}. In this paper, we are concerned primarily with individually resolvable sources or small groups of bright sources clustered in a particular direction, though our techniques may also prove useful in studies of a SB.  

When GWs from a single direction interact with the Earth, regardless of their detailed waveforms, they produce a distinctly quadrupolar correlation pattern in the timing residuals of the pulsars in a PTA.  By exploiting this fact, pulsar timing data sets from many different pulsars can be coherently combined, or ``phased-up", so as to enhance a PTA's sensitivity to GWs from particular directions. Here, we develop a formalism that enables this procedure. The strengths of this technique include:
\begin{itemize}
\item GW signals are included as part of the pulsar timing model. This means that all the issues relating to actual data sets (uneven observing cadence, different pulsars having different data spans and noise properties, the necessity to fit for pulsar parameters, etc.) are modelled simultaneously with the GW search using standard pulsar timing techniques.
\item As part of the timing model, the covariances between the GW signal and the other timing model parameters are easily obtained.
\item The algorithms underlying our technique are fast, so do not require large computing resources. The GW signal is condensed out of the raw observations into a much smaller data volume.
\item The technique makes no assumption about the actual form of the GW and so is useful for detecting unexpected GW signals as well as anticipated signals like BWMs or continuous waves (CWs) from SMBHBs in circular orbits. Inspection of the GW signal stream produced with our technique can help to determine what type of optimized signal detection schemes to implement. 
\end{itemize}

In Section 2 of this paper, we discuss how GWs manifest themselves in pulsar timing measurements and introduce our so-called $\hat{\cal A}_+$ and $\hat{\cal A}_\times$ technique. In Section 3, we describe how $\hat{\cal A}_+$ and $\hat{\cal A}_\times$ can be used in matched-filter searches for specific gravitational waveforms. In Section 4, we describe the PPTA Data Release 1 (DR1) and the simulated data sets we analyse for the remainder of the paper and indicate where these publicly-available data can be accessed. In Section 5, using $\hat{\cal A}_+$ and $\hat{\cal A}_\times$, we conduct an all-sky search for detectable GW power from any direction with any time-domain behaviour. In Section 6, we study in greater detail the lines-of-sight to the Virgo, Fornax, Norma, Perseus, and Coma galaxy clusters, nearby massive clusters that are likely origins for the first detection by PTAs of an isolated GW source; for each of these special directions, we conduct a search for BWMs, clear indicators of the final merger of an SMBHB.  In Section 7, we demonstrate that the correlations induced in PTA residuals from clock errors, inaccuracies in solar-system ephemerides, and GWs can be measured simultaneously without confusion. Finally, in Section~8, we summarize and conclude our work with some final remarks.

\section{Planar Gravitational Waves and Pulsar Timing}

A planar GW, $h_{ij}(t)$, coming from the direction ${\bf \hat{n}}$, can be expressed as
\begin{eqnarray}
h_{ij}(t)=a_+(t)\epsilon^+_{ij}+a_\times(t)\epsilon^\times_{ij},
\end{eqnarray}
where $a_+(t)$ and $a_\times(t)$ describe the time dependence of the two polarisation modes of the GW and $\epsilon^+_{ij}$ and $\epsilon^\times_{ij}$ are the relevant polarisation tensors.  Explicit representations of $\epsilon^+_{ij}$ and $\epsilon^\times_{ij}$ in ecliptic longitude and latitude are given in \citet{lwk+11}. For a PTA with $N_P$ pulsars, the apparent pulsation frequency of the $K^{\rm th}$ pulsar in direction ${\bf \hat{n}}_K$ is influenced by the GW as 
\begin{eqnarray}
\label{eq:5.2}
\frac{\Delta \nu_K}{\nu_K}&=&-\frac{1}{2}\frac{\hat{n}_K^i\hat{n}_K^j}{(1-\cos{\theta_K})}\times\nonumber\\&&\left[h_{ij}(t)|_E-h_{ij}\left(t-\frac{D_K}{c}\right)|_{{\bf r}_K}\right],
\end{eqnarray}
where $\cos{\theta_K}={\bf \hat{n}\cdot\hat{n}}_K$, the $|_E$ and $|_{{\bf r}_K}$ notation indicates that the strain field is to be evaluated at the location of the Earth and the $K^{\rm th}$ pulsar respectively, and $D_K$ is the distance from Earth to the pulsar \citep{ew75,lwk+11}. Despite the $(1-\cos{\theta_K})$ dependence in the denominator of Equation \ref{eq:5.2}, the fractional frequency shift does not diverge as $\theta_K$ approaches zero because the metric perturbation is transverse to the wave propagation direction, i.e. $\hat{n}^i_Kh_{ij}\hat{n}^j_K\sim \sin^2{\theta_K}$.  The perturbation to ToAs from the $K^{\rm th}$ pulsar caused by this GW, $\delta t_K^h(t)$, is given by the integral of the fractional frequency change in Equation~\ref{eq:5.2}:
\begin{eqnarray}
\label{eq:5.3}
\delta t_K^h(t)&=&-\frac{1}{2}\frac{\hat{n}_K^i\hat{n}_K^j}{(1-\cos{\theta_K})}\times\nonumber\\&&~~~~~~~~~~~~\left[\epsilon_{ij}^+A_+(t)+\epsilon_{ij}^\times A_\times(t)\right],\nonumber\\
&\equiv&G_K^+A_+(t)+G_K^\times A_\times(t),
\end{eqnarray}
where
\begin{eqnarray}
\label{eq:5.4}
A_{\star}(t)=\int_0^t\left[a_{\star}(t')|_E-a_{\star}(t'-D_K/c)|_{{\bf r}_K}\right]dt'.
\end{eqnarray}
The ``$\star$" subscript acts as a placeholder for either ``$+$'' or ``$\times$". The factors $G^\star_K$ depend on the relative angle between the GW source and the pulsar; they are maximized when the pulsar and GW source are in approximately the same direction of the sky and minimized when they are on opposite sides of the sky. This means that PTA sensitivity to point sources is best in the directions of the sky with the highest concentration of well-timed pulsars.

\subsection{Timing Model Development and Noise}

We can express the $n_K$ timing residuals for the $K^{\rm th}$ pulsar in vector form as
\begin{eqnarray}
\label{eq:5.5}
\delta{\bf  t}_{K} = {\bf M}_K\delta{\bf p}_K+\delta {\bf t}^h_K+\delta {\bf t}^n_K.
\end{eqnarray}
The first term on the right side of Equation \ref{eq:5.5} describes any structure in the residuals from inaccuracies in the timing model parameters.  In this linearized approximation, $\delta{\bf p}_K$, a vector describing how much the $m_K$ timing model parameters deviate from their true values, is assumed to be small. The design matrix, ${\bf M}_K$, is the $n_K\times m_K$ matrix describing how changes in the timing model parameters influence the residuals.  The second term describes any structure in the residuals induced by the gravitational wave.  The third term, $\delta {\bf t}^n_K$ describes any noise that influences the residuals. This may consist of white radiometer and pulse-phase-jitter noise, red spin noise, and a variety of additional sources \citep{cs10,sc10}. Additional correlated timing structure can be caused by things like inaccuracies in terrestrial time standards or errors in the position of the solar-system barycenter. We discuss these issues in Section 7. For our purposes here, these effects can be subsumed into the noise term of Equation~\ref{eq:5.5}, $\delta {\bf t}^n_K$.

If we temporarily neglect the influence of GWs on the timing residuals, $\delta {\bf t}^h_K$, with the noise covariance matrix ${\bf C}_K=\langle(\delta{\bf t}^n_K)\cdot(\delta{\bf t}^{n}_K)^T\rangle$, we can estimate the maximum likelihood corrections to the timing model parameters, $\delta{\bf \hat{p}}_K$, and the parameter covariance matrix, ${\bf C}^P_K$ \citep{g10}:
\begin{eqnarray}
\label{eq:5.6}
{\bf C}^P_K&=&({\bf M}_K^T{\bf C}_K^{-1}{\bf M}_K)^{-1},~{\rm and}\\
\label{eq:5.7}
\delta{\bf\hat{p}}_K&=&{\bf C}^P_K{\bf M}_K^T{\bf C}_K^{-1}\delta {\bf t}_K.
\end{eqnarray}
The use of a general noise model, first utilized by the pulsar timing community in Coles et al. (2011), is now commonly used for iteratively refining timing models as additional ToAs are acquired. \citet{chc+11} demonstrated spectral methods for adequately estimating the noise covariance matrix and showed that properly modeling the noise, especially the highly temporally correlated red spin noise seen in some pulsars, is crucial for mitigating biases in timing model parameter estimation.  However, if the residuals are being influenced by GWs, failing to account for them can lead to improper noise modeling and biased parameter estimation.  With the many PTA data sets currently available, the influence of GWs in the vicinity of Earth, which causes correlated residual structure across all timing data sets with a distinct quadrupolar pattern, can be disentangled from noise processes specific to single pulsars or errors in individual timing models.

As Equation \ref{eq:5.4} indicates, $A_\star$ is an integral of the difference in two terms: an ``Earth term" common to all pulsar timing data sets and a ``pulsar term" that differs between data sets owing to the different positions and distances of the pulsars. For burst GWs with durations much shorter than the light travel times between Earth and the pulsars in a PTA (thousands of years, typically), when the Earth term is active the various pulsar terms are all quiescent and the pulsar terms are negligible. However, the GWs from a SMBHB will be in the band of frequencies potentially detectable by PTAs for the order of millions of years. If we treat a SMBHB as a monochromatic source of sinusoidal GWs (i.e., ignore the secular frequency evolution of the source over thousand-year time scales), then each pulsar term will be a sinusoid of the same frequency as the Earth term, a nearly identical amplitude, and an essentially random and uniformly distributed phase. The pulsar term can perfectly add to the Earth term, perfectly cancel it, or anything in-between. The pulsar terms are uncorrelated in different pulsar timing data sets. If $N_P$ pulsars with approximately equal timing precision are analysed and their residuals coherently combined, the self-noise from these pulsar terms is suppressed by $N_P^{1/2}$ \citep{hd83}. For discussion of frequency evolution anticipated from SMBHBs and prospects for including pulsar terms in the signal model rather than treating them as sources of noise, see, e.g., \citet{abb+14} and \citet{wmj14}.

\subsection{Building GWs into Timing Models}

We model $A_+(t)$ and $A_\times(t)$ as linearly interpolated functions on a grid of $N_\tau$ times $\tau_\mu$.  In principle, different grids could be used for the + and $\times$ polarizations, but for simplicity, we will use an identical grid for each of them.  The grid does not need to be evenly spaced; for some pulsar timing data sets where the ToA sampling is sparse in decades-old data but relatively uniform and dense in more modern data, variable spacing in the interpolation grid may be appropriate.  An advantage to modeling $A_+(t)$ and $A_\times(t)$ as linear interpolants rather than a higher-order polynomial or Fourier series is that individual bad ToAs or small time spans of bad ToAs, either with large uncertainties or apparent biases, mainly influence the estimates of $A_+(t)$ and $A_\times(t)$ locally rather than over larger spans of data or the whole data set. 

These models for $A_+$ and $A_\times$, which we refer to as ${\cal A}_+$ and ${\cal A}_\times$, are incorporated into a global fit to the full set of ToAs that is carried out simultaneously with all the timing models for all the pulsars in the array. The ${\cal A}_+(t)$ and ${\cal A}_\times(t)$ time series must be constrained so that they are not covariant with the astrometric or spin parameters of the individual pulsar timing models. The products of this fitting procedure are $\hat{\cal A}_+$ and $\hat{\cal A}_\times$, the maximum likelihood estimators of ${\cal A}_+$ and ${\cal A}_\times$, and a matrix, ${\bf C}_{+\times}$, describing covariances in  $\hat{\cal A}_+$ and $\hat{\cal A}_\times$. 

The $\hat{\cal A}_+$ and $\hat{\cal A}_\times$ time series are a complete representation of the GWs coming from a particular direction in that they carry all the information with respect to such GWs that is contained in the ToAs. One cannot obtain a better signal estimator using the ToAs than can be done with the auxiliary time series alone. Full details of the implementation of the global fit and the constraints that need to be applied to ${\cal A}_+(t)$ and ${\cal A}_\times(t)$ during the fitting procedure are given in Appendix~A. These algorithms have been implemented in the \textsc{tempo2} software package \citep[][see Appendix B for usage details]{ehm06}.

\section{Searching $\hat{\cal A}_+$ and $\hat{\cal A}_\times$ for Anticipated GW Waveforms}

Although one of the great strengths of the $\hat{\cal A}_+$ and $\hat{\cal A}_\times$ time series is that it can be used to study GWs without an underlying signal model, we demonstrate here how searches for specific waveforms can be implemented directly in the $\hat{\cal A}_+$ and $\hat{\cal A}_\times$ time series after they have been computed. \citet{zhw+14} developed a procedure for conducting a matched-filter search for a specific waveform in $\hat{\cal A}_+$ and $\hat{\cal A}_\times$ and applied it to Data Release 1 (DR1) from the PPTA in a search for CWs. The upper limit on the amplitude of CWs derived in that paper is among the most stringent achieved to date \citep[for competitive limits, see][]{abb+14,bps+15}. Here, we will review the essential elements needed to carry out a matched-filter search in $\hat{\cal A}_+$ and $\hat{\cal A}_\times$ and elaborate on certain points, but will focus on a different, simpler signal model: a BWM. We will apply these methods in a small targeted search for BWMs from nearby massive galaxy clusters using DR1 later in this paper. \citet{whc+15} searched DR1 for BWMs but did not utilize $\hat{\cal A}_+$ and $\hat{\cal A}_\times$ time series. Our search is different and complementary to the work described in that paper; $\hat{\cal A}_+$ and $\hat{\cal A}_\times$ techniques produce consistent results with \citet{whc+15} and are ultimately more computationally efficient.

In conducting the global fit that generates $\hat{{\cal A}}_+$ and $\hat{{\cal A}}_\times$, we must compute a global parameter covariance matrix (see Appendix A for a detailed discussion). A $2N_\tau\times 2N_\tau$ sub-block of this matrix describes the correlated uncertainties in $\hat{{\cal A}}_+$ and $\hat{{\cal A}}_\times$; we call this sub-block $\bf{C}_{+\times}$. As discussed in \citet{zhw+14}, ${\bf C}_{+\times}$ is not a full-rank matrix because of the constraints applied to $\hat{{\cal A}}_+$ and $\hat{{\cal A}}_\times$ and is thus non-invertible. We define $\tilde{{\bf C}}_{+\times}={\bf EFE}^T$ where $\bf{E}$ is a matrix containing the eigenvectors of ${\bf C}_{+\times}$ associated with non-null eigenvalues and $\bf{F}$ is a square diagonal matrix containing those non-null eigenvalues. This generalized covariance matrix is the covariance matrix on the subspace orthogonal to the constraints and is manifestly invertible: ${\bf \tilde{C}}_{+\times}^{-1}={\bf EF}^{-1}{\bf E}^T$. Also, we define the projection operator ${\bf P}={\bf EE}^T$. 

We define a vector $\hat{{\cal A}}^T=[\hat{{\cal A}}_+^T,\hat{{\cal A}}_\times^T]$, i.e., $\hat{{\cal A}}_+$ and $\hat{{\cal A}}_\times$ stacked into a single vector. If we assume that the structure in $\hat{{\cal A}}$ is due to a BWM occurring at $t_{\rm inj}$, if constraints had not been applied, $\hat{{\cal A}}$ could be expressed as a linear combination of two basis elements:
\begin{eqnarray}
\beta_+&=&\left[\begin{array}{c}({\bf \tau}-t_{\rm inj})\Theta({\bf \tau}-t_{\rm inj})\\0\cdot{\bf\tau} \end{array}\right],\\\beta_\times&=&\left[\begin{array}{c}0\cdot{\bf \tau}\\({\bf \tau}-t_{\rm inj})\Theta({\bf \tau}-t_{\rm inj})\end{array}\right],
\end{eqnarray}
where ${\bf \tau}$ is the vector of $N_\tau$ times at which the $\hat{\cal A}_+$ and $\hat{\cal A}_\times$ time series are sampled and $\Theta$ is the Heaviside step function. This signal model for a BWM is discussed by, e.g., \citet{pbp10}, \citet{vl10}, \citet{cj12}, and \citet{mcc14}. The appropriately constrained version of $\hat{{\cal A}}$ can be expressed as a linear combination of these basis elements after they have been projected into the subspace orthogonal to the constraints, i.e., as a linear combination of $\tilde{\beta}_+={\bf P}\beta_+$ and $\tilde{\beta}_\times={\bf P}\beta_\times$. We can say 
\begin{eqnarray}
\hat{{\cal A}}=\alpha_+\tilde{\beta}_++\alpha_\times\tilde{\beta}_\times, 
\end{eqnarray}
and we can compute maximum likelihood estimates of the coefficients, $\hat{\alpha}_+$ and $\hat{\alpha}_\times$, as follows:
\begin{eqnarray}
\hat{\alpha}=\left({\bf B}^T{\bf \tilde{C}}_{+\times}^{-1}{\bf B}\right)^{-1}{\bf B}^T{\bf \tilde{C}}_{+\times}^{-1}\hat{{\cal A}},
\end{eqnarray}
where $\alpha^T=[\alpha_+,\alpha_\times]$ and ${\bf B}=[\tilde{\beta}_+,\tilde{\beta}_\times]$. The uncertainties on the estimates $\hat{\alpha}_+$ and $\hat{\alpha}_\times$ are encoded in the $2\times2$ matrix 
\begin{eqnarray}
{\bf\Sigma}=\left({\bf B}^T{\bf \tilde{C}}_{+\times}^{-1}{\bf B}\right)^{-1}. 
\end{eqnarray}

This signal parameter estimator, $\hat{\alpha}$, is biased. The bias is a consequence of the constraints applied to $\hat{\cal A}_+$ and $\hat{\cal A}_\times$. As an extreme example, if a GW signal was purely quadratic, that signal would be entirely filtered out of $\hat{\cal A}_+$ and $\hat{\cal A}_\times$ by the constraints and the GW signal could never be reconstructed. \citet{brn84} and \citet{mcc13} describe the procedure of fitting certain functional forms out of timing residuals as equivalent to applying a frequency-domain filter to the residual fluctuation spectrum. Analogously, the constraints applied to $\hat{\cal A}_+$ and $\hat{\cal A}_\times$ destructively filter out some features of the GW signal. However, the spectral shape of constraint-induced filters depends upon the timespan of the data set. As more data are gathered, the constraints and the GW signal become less covariant, less of the GW signal is filtered out, and the bias in estimation of the signal parameters is reduced. Furthermore, through simulations, the bias can be quantified, allowing for robust and trustworthy parameter estimation.

\section{Real and simulated data sets}

For much of the remainder of this paper, we work with both simulated and real data sets. Our simulated data sets were produced using a newly developed software package called \textsc{ptaSimulate}. The simulated data sets we analyse are\footnote{All the data sets (real and simulated) used in this paper are available from  \href{http://dx.doi.org/10.4225/08/560A00E2036F6}{http://dx.doi.org/10.4225/08/560A00E2036F6}}:

\begin{itemize}
\item \textsc{Sim1}: 20 synthetic pulsars are generated with random sky positions (statistically uniform on the sphere). We assume weekly ToA measurements with 50 ns timing precision. The timing models for each pulsar only include fits for three spin parameters (rotational phase, frequency, and frequency derivative). The timing residuals for each pulsar are consistent with 50 ns rms white Gaussian noise.

\item \textsc{Sim2}: As in \textsc{Sim1}, but a bright GW burst has been injected into both GW polarisation channels of the simulation from a source at right ascension 0 h and declination 0$^\circ$. The pulsar locations differ from those in \textsc{Sim1}, but are again drawn from a distribution that is uniform on the sphere. The white Gaussian noise added to each ToA again has an rms of 50 ns, but we use a different realisation of noise from that in \textsc{Sim1}.

\item \textsc{Sim3}: We have generated 20 simulated pulsars with the same positions as the 20 pulsars comprising the PPTA DR1. We have assumed these pulsars are observed every two weeks and have ToA uncertainties consistent with 100 ns rms Gaussian white noise. In each pulsar timing model, we fit only for rotational phase, frequency, and frequency derivative. We have simulated a clock error by generating the simulated data with TAI and reprocessing them with BIPM2013, two different realizations of terrestrial time. We have simulated an ephemeris error by using JPL ephemeris DE414 \citep{s06} to generate the simulated data and DE421 \citep{fwb08} to reprocess it. 

\item \textsc{Sim4}: As in \textsc{Sim3}, but a GW burst has also been injected into the simulated timing data. The realisation of 100 ns rms Gaussian white noise in \textsc{Sim4} differs from that in \textsc{Sim3}.
\end{itemize}

We analyse a version of the PPTA DR1 \citep{mhb+13} that has been updated to include detailed noise models developed and used by \citet{whc+15} and \citet{zhw+14}.  The noise models account for additional white noise in the data beyond that anticipated from radiometer noise alone (jitter is a likely contribution to this) and, using the techniques developed in \citet{chc+11}, correlated red timing noise.

\section{All-sky Searches for Excess GW Power}

\begin{figure*}
\label{fig:1}
\begin{center}
\includegraphics[scale=0.36]{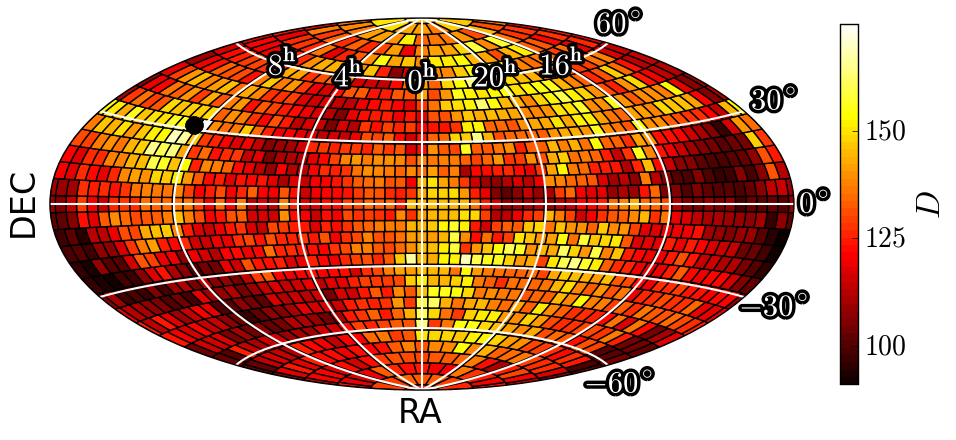}
\includegraphics[scale=0.29]{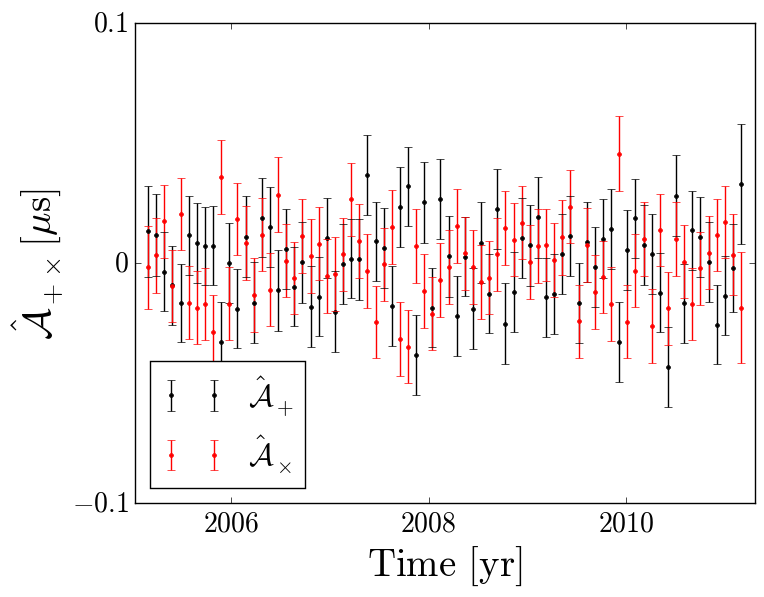}\\
\includegraphics[scale=0.36]{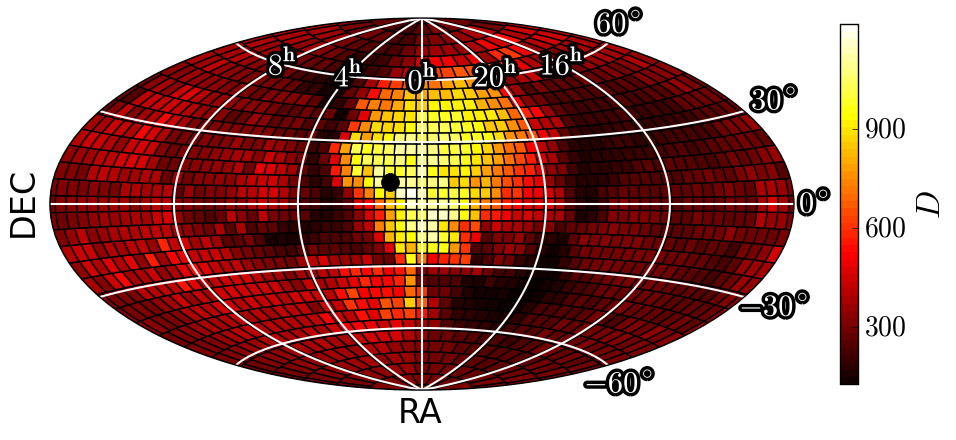}
\includegraphics[scale=0.29]{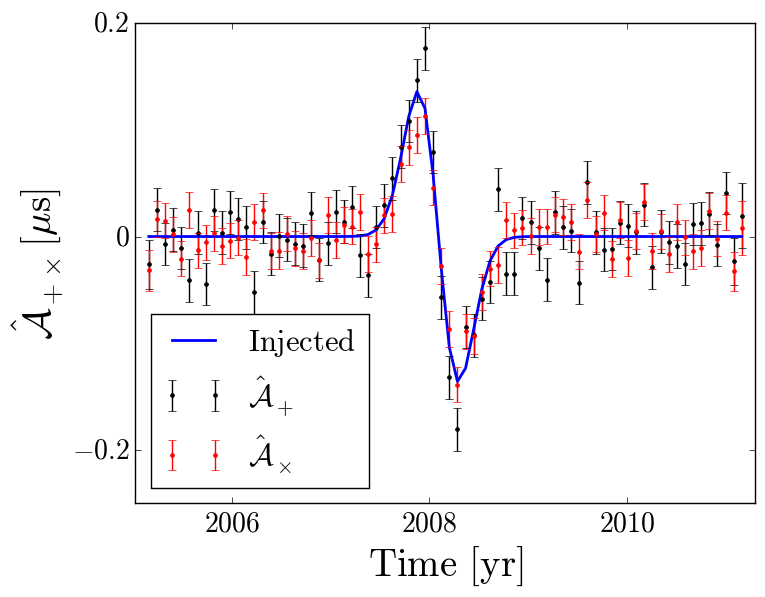}\\
\includegraphics[scale=0.36]{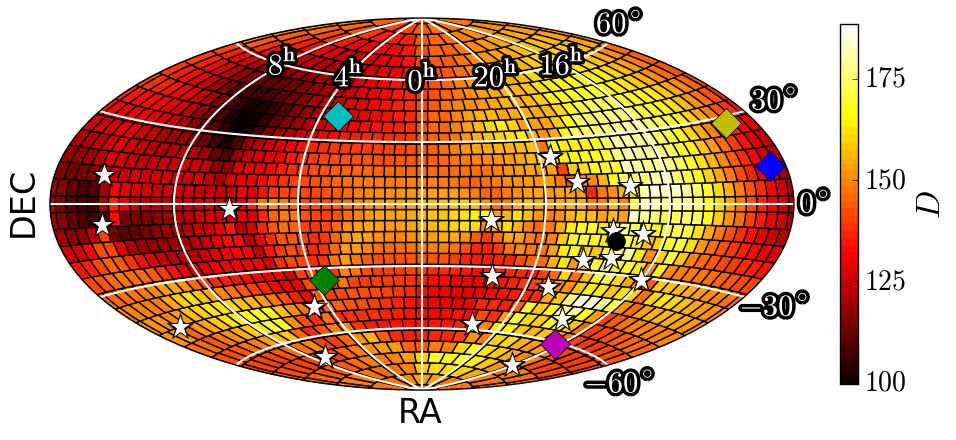}
\includegraphics[scale=0.29]{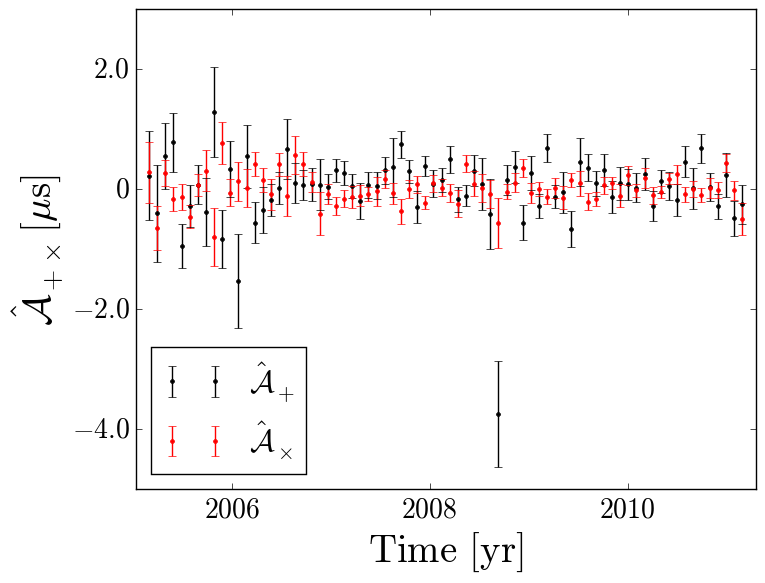}
\end{center}
\caption{{\bf Left}: the total power detection statistic, $D$, as it varies over the sky. The black circle indicates the position on the sky that yielded the greatest value of $D$. {\bf Right}: the $\hat{\cal A}_+$ and $\hat{\cal A}_\times$ time series for the direction of the sky that yielded the greatest value of $D$. {\bf Top}: Analysis of \textsc{Sim1}. The detection statistic never exceeds values that are inconsistent with noise.  {\bf Middle}: Analysis of \textsc{Sim2}. The detection statistic clearly indicates that the data is inconsistent with the noise, we are able to localise the location of the injected GW source, and we can reconstruct the waveform of the injected burst. {\bf Bottom}: Analysis of the six-year Data Release 1 (DR1) from the Parkes Pulsar Timing Array. The stars indicate the positions of the 20 pulsars included in Data Release 1. The coloured diamonds indicate the positions of five nearby massive galaxy clusters: Virgo (blue), Fornax (green), Norma (magenta), Perseus (cyan), and Coma (yellow). There is no significant evidence for GW power in DR1.}
\end{figure*}

Here, we make use of the $\hat{\cal A}_+$ and $\hat{\cal A}_\times$ time series to search for GW power coming from any direction on the sky. This search would detect sufficiently strong signals of any waveform. As a detection statistic for this search, we define a dimensionless measure of the total GW power contained in $\hat{\cal A}_+$ and $\hat{\cal A}_\times$:
\begin{eqnarray}
D=\hat{\cal A}^T{\bf \tilde{C}}_{+\times}^{-1}\hat{\cal A}.
\end{eqnarray}
This is the mean squared error statistic for the null hypothesis. Using slightly different terminology, $D=\hat{\cal A}^T_W\hat{\cal A}_W$ where $\hat{\cal A}_W={\bf U}^{-1}\hat{\cal A}$ and ${\bf U}$ is a Cholesky whitening matrix, i.e. $\tilde{\bf C}_{+\times}={\bf UU}^T$ \citep{chc+11}. In the absence of signal, assuming the noise is correctly modeled by ${\bf \tilde{C}}_{+\times}$ and thus ${\bf U}$, the whitened data stream, $\hat{\cal A}_W$, will consist of Gaussian noise, be unit variance, and white; $D$ will follow $\chi^2$ statistics with $N_{\rm dof}=2(N_\tau-N_c)$ degrees of freedom, where $N_c$ is the number of constraints applied individually to the $\hat{\cal A}_+$ and $\hat{\cal A}_\times$ time series and, again, $N_\tau$ is the number of grid points at which $\hat{\cal A}_+$ and $\hat{\cal A}_\times$ are sampled. If $\hat{\cal A}_W$ is not unit variance and white, either GW signal is present in the data or the individual pulsar noise models are inaccurate. The noise models for the data we use here have been vetted by the analyses of \citet{zhw+14} and \citet{whc+15}.

We compute $\hat{\cal A}_+$ and $\hat{\cal A}_\times$ over a grid of trial source locations on the sky and compute $D$ at each grid point. The number of statistically independent samples on this grid of sky positions, $N_s$, influences our anticipated false alarm probability. The false alarm probability for a given value of $D$ is
\begin{eqnarray}
f(D)&=&1-c(D;N_{\rm dof})^{N_s},
\end{eqnarray} 
where $c(D;N_{\rm dof})$ is the cumulative distribution for a $\chi^2$ distribution with $N_{\rm dof}$ degrees of freedom evaluated at $D$. \citet{cv14} recently demonstrated that the response of a PTA to any type of GW can be expressed as a linear combination of $N_P$ orthogonal modes, or sky maps, where, again, $N_P$ is the number of pulsars in the array. A similar result was derived by \citet{grt+14}. Since our actual data consists of observations of 20 pulsars, we will use the result from these authors and conservatively set $N_s=20$.

In Figure 1, we display the results of this search procedure when applied to \textsc{Sim1}, \textsc{Sim2}, and DR1 (listed in order from top to botton). We have sampled $\hat{\cal A}_+$ and $\hat{\cal A}_\times$ at 74 evenly spaced epochs, corresponding to a cadence of 30 days. The left panels indicate the value of $D$ at each sky position. The black dot in these left panels indicates the position on the sky that yielded the greatest value of $D$. The right panels depict the $\hat{\cal A}_+$ and $\hat{\cal A}_\times$ time series associated with the direction on the sky that yielded this greatest value of the detection statistic. For the analysis of \textsc{Sim1}, our null example, $\hat{\cal A}_+$ and $\hat{\cal A}_\times$ appear by eye to be consistent with zero signal and, indeed, the maximum value of $D$ that is realised is 174.8, consistent with approximately 18.7\% of realizations of noise.

Our analysis of \textsc{Sim2} is shown in the middle row of Figure 1. The position of the GW source was set to a right ascension of 0 h and a declination of $0^\circ$. Notice the change in the colour scales between the top left and middle left panels. In the middle row, the detection statistic at nearly every trial sky position is inconsistent with noise, but $D$ is peaked around the location of the simulated source demonstrating the efficacy of this procedure in localising a bright GW source on the sky. The $\hat{\cal A}_+$ and $\hat{\cal A}_\times$ time series produced at the position of the maximal detection statistic reproduces the injected GW signal,
\begin{eqnarray}
b(t)=b_0(t-t_*)e^{-(t-t_*)^2/2w^2},
\end{eqnarray}
where $b_0=3\times10^{-9}$, $t_*={\rm MJD}~54500$, and $w=75~{\rm days}$. For simplicity, we injected the same functional form into each polarisation channel. This burst, at maximum amplitude, produces an approximately 136 ns ToA perturbation. In our analysis of \textsc{Sim1} and \textsc{Sim2}, we have shown that with idealised data, our search method correctly returns a null result in the absence of a GW signal, but does correctly detect, reproduce, and localise on the sky, a bright signal that is present.

\begin{figure*}
\label{fig:2}
\begin{center}
\includegraphics[scale=0.51]{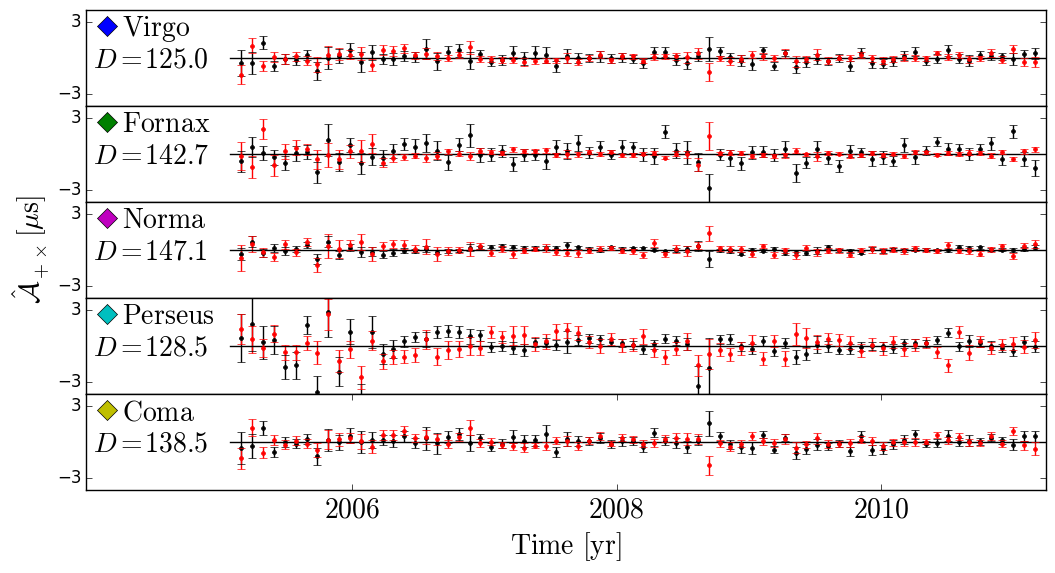}
\end{center}
\caption{The $\hat{\cal A}_+$ (black) and $\hat{\cal A}_\times$ (red) time series produced when DR1 is phased up to the locations of the massive Virgo, Fornax, Norma, Perseus, and Coma galaxy clusters. All pointings are consistent with noise only (as indicated by the $D$ value quoted in each panel). The diamonds in each panel are colour coded with the diamonds in the bottom left panel of Figure 1 indicating the directions to these clusters.}
\end{figure*}

The bottom row of plots in Figure 1 depict the results of our search for directional GWs in the PPTA's DR1.  The search is carried out in the same way as for the simulated data sets, but we have used a generalised least-squares fitting routine \citep[as described in][]{chc+11} in order to account for the different red noise present in different pulsars\footnote{In contrast to searches for a GW background, the signals of interest here have significantly different spectral characteristics than the intrinsic red noise often observed in pulsar data sets.}. The white stars in the bottom left panel indicate the positions of the 20 pulsars comprising DR1. The distribution of pulsars on the sky does not approach statistical uniformity and the noise properties of the pulsars are not all equal as in \textsc{Sim1} and \textsc{Sim2}. With $N_\tau=74$, $N_c=7$ (as discussed in Appendix~A), and $N_s=20$, $D$ would have to exceed approximately 194 to claim an excess of GW power with 99\% confidence or 184 for 95\% confidence. The maximum value of $D$ we find in our analysis of DR1 is 188.5, which is between the 95\% and 99\% confidence thresholds. 

Even though this value of $D$ is consistent with a null detection, it is close to values of interest. However, there is a reason to be skeptical of this marginally high value of $D$. The $\hat{\cal A}_+$ time series associated with this maximal value of $D$ from DR1 has a $4.2\sigma$ outlier at an epoch near September of 2008. This epoch is contemporaneous with the commissioning of a new pulsar timing backend at Parkes. Many ToAs from this brief era have been excluded from DR1 for displaying obvious systematic issues, but it is likely that systematic artefacts may still be present in the data. If this one outlier in the $\hat{\cal A}_+$ time series is artificially forced towards zero until its $1\sigma$ error bar is consistent with zero, the value of $D$ from this pointing is reduced to approximately 171, consistent with approximately 30\% of noise realisations. 

This result with DR1 underscores the importance of the IPTA project in two ways. Firstly, if the marginally significant result we have discussed is indeed just a systematic artefact in PPTA data, comparison with measurements from the EPTA and NANOGrav from around September of 2008 of a subset of pulsars that overlaps with the PPTA sample should be able to resolve the issue. Secondly, if the outlier in the PPTA data set is a very short duration, very bright GW burst, it will be present in the data sets of all three PTAs comprising the IPTA and a joint analysis of the combined data sets will bolster the significance of the result.

\section{Targeted Investigations of Nearby Massive Galaxy Clusters}

\citet{spl+14} recently conducted a detailed analysis of surveys of local galaxies in order to identify potential directions on the sky from which an initial detection of resolvable GWs by PTAs is likely to originate, so-called ``GW hotspots". They singled out, in order of increasing distance from Earth, several massive galaxy clusters: Virgo, Fornax, Norma, Perseus, and Coma. The directions to these clusters are indicated in the bottom left panel of Figure 1 with colored diamonds. These clusters are all within 100 Mpc of Earth and all but Fornax, the smallest of the five, contain upwards of 500 galaxies. In Figure 2, we plot the $\hat{\cal A}_+$ and $\hat{\cal A}_\times$ time series produced when DR1 is phased up to the directions of each of these galaxy clusters. All are consistent with noise.

There are several notable features in Figure 2. First, all five pointings have some marginal outliers near the September 2008 commissioning of the new pulsar timing backend at Parkes that was discussed earlier. We demonstrated in our analysis of \textsc{Sim2} (depicted in Figure 2) that a GW burst can be localised on the sky and the five clusters we consider here are from widely separated directions; this gives further weight to the argument that there are systematic issues with DR1 near September of 2008 and it is very unlikely that the excess power in $\hat{\cal A}_+$ and $\hat{\cal A}_\times$ is from a spatially localised GW source. Second, the scatter in all pointings is larger earlier in the time span; this is due largely to an increase in the observing cadence over time and thus a greater number of ToAs contributing to each $\hat{\cal A}_+$ and $\hat{\cal A}_\times$ sample later in the data set. The change in the observing cadence has not led to any issues in our analysis of DR1, but this illustrates the possible need for unequal spacing in the $\hat{\cal A}_+$ and $\hat{\cal A}_\times$ sample grid in other data sets. Finally, the scatter in the $\hat{\cal A}_+$ and $\hat{\cal A}_\times$ time series is biggest in our pointing towards the Perseus cluster and smallest in our pointing towards the Norma cluster. The Perseus cluster is in the opposite direction of the sky from the peak concentration of PPTA pulsars while the Norma cluster is very nearly in the same direction as many of the PPTA pulsars. This is a reflection of the point we discussed in Section 2 that pulsar timing measurements are more sensitive to GWs coming from directions of the sky near the direction of the pulsar. It is well known that an anisotropic distribution of well-timed pulsars leads to anisotropic sensitivity to GWs; this again highlights the importance of the IPTA and close collaboration and data sharing between pulsar astronomers in the Northern and Southern Hemispheres.

 \begin{figure}
 \label{fig:3}
\begin{center}
\includegraphics[scale=0.42]{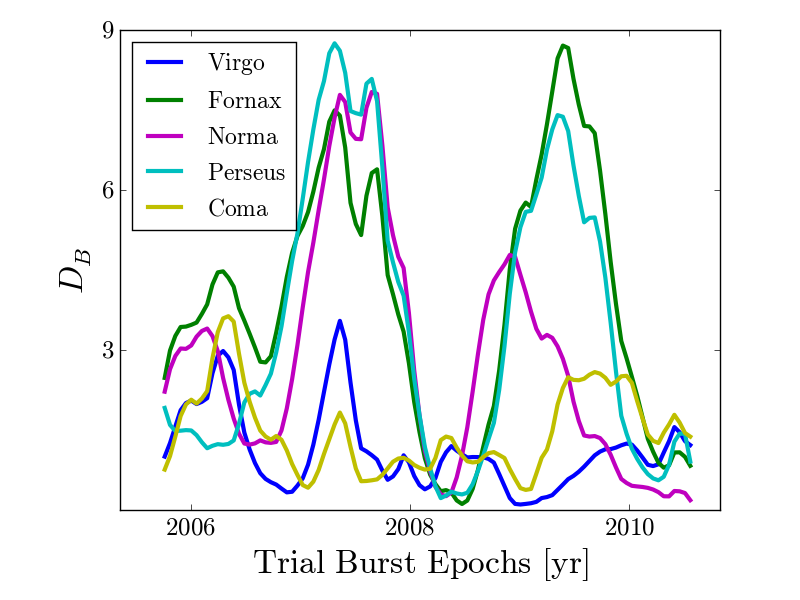}\\
\includegraphics[scale=0.42]{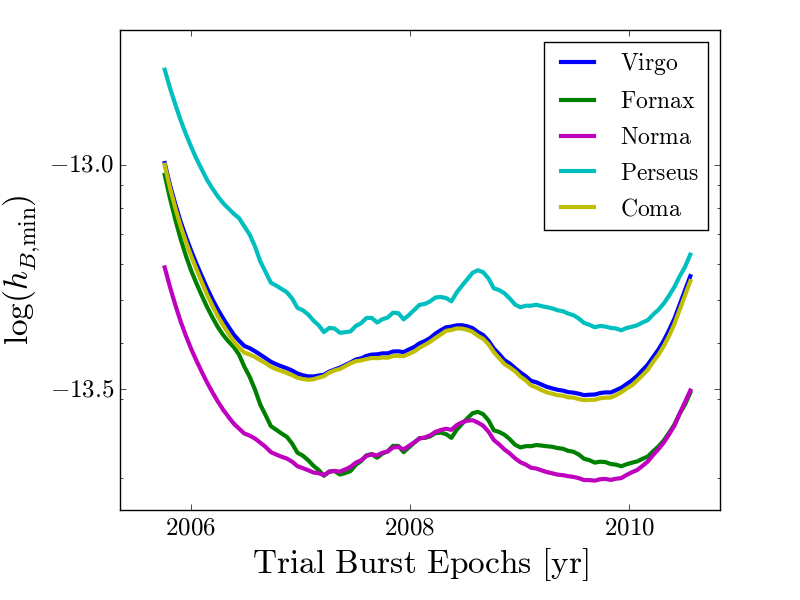}\\
\includegraphics[scale=0.42]{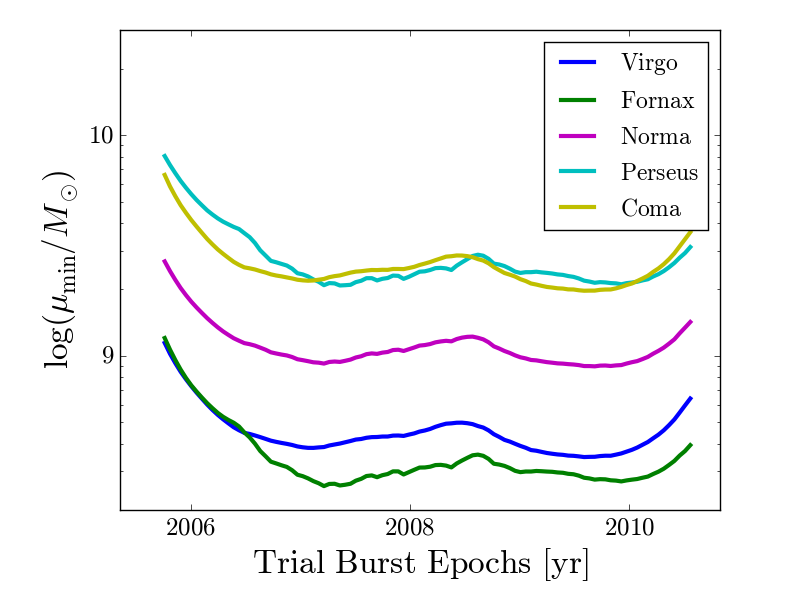}
\end{center}
\caption{{\bf Top}: the detection statistic, $D_B$, derived from a search for BWMs in the $\hat{\cal A}_+$ and $\hat{\cal A}_\times$ time series associated with the directions to five nearby massive galaxy clusters. $D_B$ needs to exceed approximately 9.2 in order to be inconsistent with more than 95\% of noise realizations. {\bf Middle}: the minimum amplitude of a BWM that would have yielded a value of $D_B$ exceeding 9.2. {\bf Bottom}: Assuming a statistically average inclination angle for a merging SMBHB, based on the luminosity distance to each of the five galaxy clusters we investigate, the minimum reduced mass of a merging SMBHB that would have produced a detectable BWM.}
\end{figure}

\subsection{A BWM Search in $\hat{\cal A}_+$ and $\hat{\cal A}_\times$}
Although we find no evidence for excess GW power in the directions of any of these five clusters, to search for a specific waveform, a matched-filter search of $\hat{\cal A}_+$ and $\hat{\cal A}_\times$ is more sensitive than a total power search. Building on our discussion from Section 3, as a test case, we here conduct a search for BWMs in these five pointings. 

We define a different detection statistic for this search: $D_B=\hat{\alpha}^T{\bf \Sigma}^{-1}\hat{\alpha}$. This is nearly identical to the detection statistic used in the BWM search conducted by \citet{whc+15}, but here we do our search directly in $\hat{\cal A}_+$ and $\hat{\cal A}_\times$. In the absence of signal, $D_B$ will follow $\chi^2$ statistics with 2 degrees of freedom. We will search epochs within the innermost 80\% of each of the five $\hat{\cal A}_+$ and $\hat{\cal A}_\times$ pointings in Figure~2 for evidence of a BWM occurring. We restrict ourselves to this window because detecting a BWM requires the ability to accurately assess the pulsar timing behaviour both pre- and post-burst. \citet{abb+15} recently showed that in a BWM search over many trial burst epochs, there are approximately 5 statistically independent trials; this fact must be accounted for in assessing the false alarm probability in our search. In order for the data to be inconsistent with 95\% of realizations of noise, with $N_s=5$, $D_B$ must exceed approximately 9.2. 

In the top panel of Figure 3, we show the values of $D_B$ derived from our search; we find nothing inconsistent with noise. Knowing ${\bf \Sigma}$, however, allows us to compute the minimum value of $|\alpha|$ needed to exceed the 95\% confidence threshold in $D_B$. We display this quantity, which we call $h_{B,{\rm min}}$, in the middle panel of Figure 3. The strain amplitude of a BWM is $h_B\approx1.5\times10^{-13}(\mu/10^9M_\odot)(d/10 {\rm Mpc})^{-1}$ where $\mu$ is the reduced mass of the binary and $d$ is the luminosity distance between the binary and Earth; we have assumed the binary has a typical inclination angle of $\pi/3$ \citep{mcc14}. Taking the luminosity distances to Virgo, Fornax, Norma, Perseus, and Coma, as 17~Mpc, 19~Mpc, 68~Mpc, 74~Mpc, and 99~Mpc, respectively \citep[see][and references therein]{spl+14}, we have generated the bottom panel of Figure 3 displaying the minimal reduced mass, $\mu_{\rm min}$, of a merging SMBHB that would have produced a BWM bright enough to exceed our 95\% confidence threshold on $D_B$. The likelihood of such a massive merger occurring in one of these galaxy clusters during our observing span is exceedingly small, but as the span of our data set grows, our sensitivity to BWMs will improve and we will become sensitive to less massive mergers.

The $\hat{\cal A}_+$ and $\hat{\cal A}_\times$ time series associated with the directions towards these five galaxy clusters can be accessed along with DR1 and our simulated data sets following the link mentioned above. Instructions by which results for any other pointing can be quickly produced with \textsc{tempo2} can be found in the usage details of Appendix B. Following the BWM example detailed here and the analysis of \citet{zhw+14}, analogous searches for any type of parameterised waveform can be easily carried out.

\section{Multipole Menagerie: Clock Errors, Inaccurate Ephemerides, and Gravitational Waves}

\begin{figure*}
\label{fig:4}
\begin{center}
\includegraphics[scale=0.4]{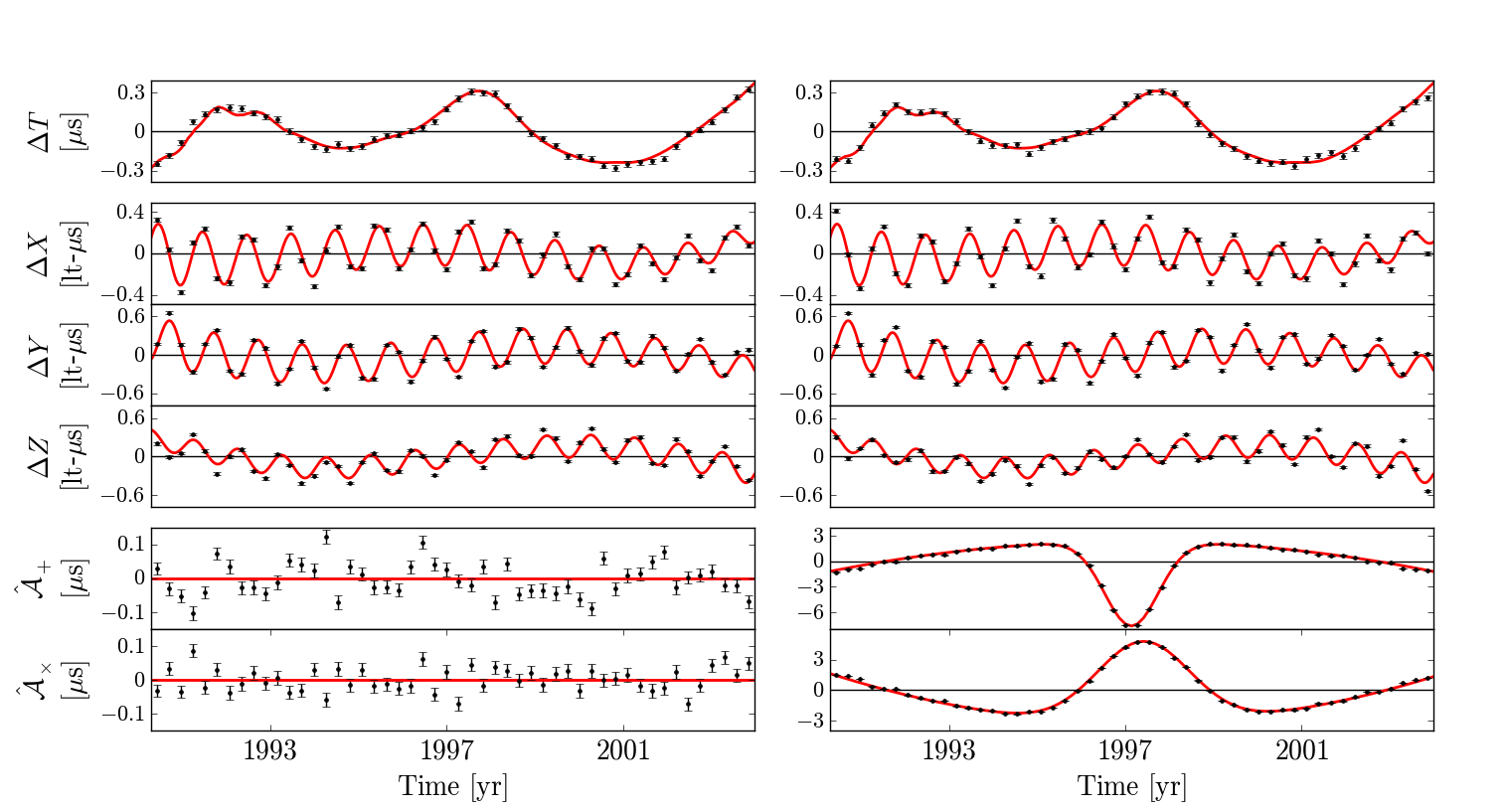}
\end{center}
\caption{Simultaneous extraction from a twenty-pulsar simulated PTA of the monopolar signature of clock errors (top panel), the dipolar signature of an inaccurate solar-system ephemeris (the second, third, and fourth panels from the top), and the two polarization modes of a GW (the bottom two panels). The left panels, depicting an analysis of \textsc{Sim3}, contain no GW signal, while the right panels, depicting an analysis of \textsc{Sim4} contain the bright GW burst described in Equation \ref{eq:burst}.}
\end{figure*}

As we mentioned in Section 1, GWs are not the sole means by which timing residuals from all the pulsars in a PTA can become correlated. Faults in terrestrial time standards can induce monopolar correlations between pulsar timing data sets \citep{hcm+12}. Errors in our estimated position for the solar-system barycenter (SSB) from, for example, inaccuracies in the measured mass of Saturn can lead to dipolar correlations between pulsar timing residuals \citep{chm+10}. In order for $\hat{\cal A}_+$ and $\hat{\cal A}_\times$ time series to be useful in studies of GW-induced inter-pulsar timing correlations, they must be able to reliably differentiate the distinctly quadrupolar signature of a GW from the monopolar or dipolar signatures of these other effects; in Figure 4, we demonstrate that our techniques can effectively do this.

The left panels of Figure 4 depict our analysis of \textsc{Sim3} in which clock errors and ephemeris errors have been simulated but there is no GW signal present. We simultaneously fit for the timing models of all the pulsars in our simulated array along with linear-interpolant models for clock errors \citep[as in][]{hcm+12}, vectorial offsets between the true SSB and its assumed location \citep[as in][]{chm+10}, and the $\hat{\cal A}_+$ and $\hat{\cal A}_\times$ time series for the direction on the sky from which we eventually inject a GW burst in \textsc{Sim4}. From top to bottom, we display the clock signal (we call this $\Delta T$), the three Cartesian components of the simulated vector displacement in the assumed SSB (called $\Delta X$, $\Delta Y$, and $\Delta Z$), and the $\hat{\cal A}_+$ and $\hat{\cal A}_\times$ time series. The red lines indicate the injected signal with the best-fit (in a least-squares sense) quadratic removed; the quadratic removal is necessary because of the quadratic fit carried out for each pulsar. The black dots indicate the recovered signal. We are able to accurately recover the clock and ephemeris-error signals present in the data and the $\hat{\cal A}_+$ and $\hat{\cal A}_\times$ are consistent with a null result. Significant monopolar and dipolar correlations present in the data have not produced spurious signal in the $\hat{\cal A}_+$ and $\hat{\cal A}_\times$ channels.

In the right panels of Figure 4 we have carried out the same procedure described above but applied to \textsc{Sim4} which contains a simulated GW burst from RA = 0.2 rad and DEC = -0.3 rad, a place on the sky that is remote from the core concentration of PPTA pulsars. The waveforms injected into the two GW polarisation channels are
\begin{eqnarray}
\label{eq:burst}
F_\star(t)={\cal F}_\star e^{-(t-t_\star)^2/2w_\star^2},
\end{eqnarray}
where ${\cal F}_+=10^{-5}$, ${\cal F}_\times=-8\times 10^{-6}$, $t_+=$ MJD 50500, $t_\times=$ MJD 50600, $w_+=224$ days, and $w_\times = 387$ days, The injected clock and ephemeris error signals are identical between \textsc{Sim3} and \textsc{Sim4}. The injected GW signal is successfully recovered, and even in the presence of a bright GW signal, the clock- and ephemeris-error signals are also successfully recovered. 

The examples depicted in Figure 4 illustrate the ability to successfully differentiate the correlations induced in pulsar timing data sets from clock errors, ephemeris errors, and GWs and to recover the relevant signals without substantial bias. However, suppose someone was only interested in searching for a GW signal. If a clock or ephemeris error were leading to correlations across multiple pulsar timing data sets and only $\hat{\cal A}_+$ and $\hat{\cal A}_\times$ were fitted, the unaccounted-for correlations from the clock or ephemeris error would bias the GW signal-recovery and potentially lead to spurious claims of GW detection \citep[for related discussion, see ][]{thk+15}. We emphasize the importance of accounting for potential clock- or ephemeris-error correlations in any PTA effort to detect GWs.

\section{Conclusions}

In this paper, we have described a technique that allows the signal of a planar GW from a particular direction of the sky to be distilled from PTA data. This plane wave could be from a single distant source, such as an individual SMBHB, or from a collection of distant sources that are sufficiently close together on the sky like a massive galaxy cluster; how close together sources need to be depends on the angular resolution of the PTA which depends on the spatial distribution of well-timed pulsars. The angular resolution of PTAs will improve over time as timing techniques are improved and as new pulsars are discovered and incorporated into PTAs.

We have assumed that GWs behave according to the predictions of linearized general relativity. Some alternative theories of gravity allow for additional polarisation modes of GWs such as longitudinal or breathing modes \citep{l13}; our formalism cannot currently address these alternative theories of gravity, but could with modification. With these caveats regarding the polarisation of GWs in mind, since our methods require no assumptions about the time-domain behaviour of GWs, they provide a general and flexible tool that will help maximize the discovery potential of PTAs, potentially facilitating the discovery of wholly unanticipated sources.

Most modern efforts to detect GWs with PTAs utilize sophisticated Bayesian inference codes \citep{vlj+11,vv14,abb+14,ltm+15}. Our total-power all-sky GW search in Section 5 and our targeted search for BWMs in Section 6 are frequentist in nature, but as $\hat{\cal A}_+$ and $\hat{\cal A}_\times$ are built into the PTA timing model, they can be straightforwardly incorporated into Bayesian analyses. We feel that the $\hat{\cal A}_+$ and $\hat{\cal A}_\times$ method will complement Bayesian GW searches by providing important diagnostic time series that will allow by-eye examination of GW signals. Furthermore, $\hat{\cal A}_+$ and $\hat{\cal A}_\times$ time series facilitate a significant condensation of the data volume necessary for GW investigations, and as such, can potentially reduce the computational resources required for even very advanced Bayesian search codes.

Finally, we point out that modern PTA data sets are constantly growing and evolving.  As technology advances and allows for upgrades in radio instrumentation and as more nations and observatories become major contributors to PTA endeavors, the data comes in a wider variety of forms and from more varied sources. PTA data are becoming more and more heterogeneous and unwieldy to work with. For the purposes of GW investigations, $\hat{\cal A}_+$ and $\hat{\cal A}_\times$ time series will aid in the creation of a compressed and homogenized auxiliary data set that isolates all of the important GW information.

\vspace{0.5cm}
For this work, we have used data from the Parkes radio telescope. The Parkes radio telescope is part of the Australia Telescope, which is funded by the Commonwealth of Australia for operation as a National Facility managed by the Commonwealth Scientific and Industrial Research Organisation (CSIRO). DRM is a Jansky Fellow of the National Radio Astronomy Observatory (NRAO). NRAO is a facility of the National Science Foundation (NSF) operated under cooperative agreement by Associated Universities, Inc. DRM also acknowledges support from a sub-award to Cornell University from West Virginia University through NSF/PIRE grant 0968296, the NANOGrav Physics Frontiers Center NSF award PHY-1430284, and the New York NASA Space Grant Consortium. X-JZ and LW acknowledge funding support from the Australian Research Council. GH is supported by Australian Research Council grant FT120100595. JW is supported by NSFC project No. 11403086 and the West Light Foundation CAS XBBS201322. SO is supported by the Alexander von Humboldt Foundation.

\appendix
\section{Approximate Representation of $A_+$ and $A_\times$}

The interpolated approximations to $A_+$ and $A_\times$ that we make use of can be written as
\begin{eqnarray}
{\cal A}_{\star}(t)&=&\sum_{\mu=1}^{N_\tau-1}l_{\star,\mu}(t)w(t,\tau_\mu,\tau_{\mu+1}),
\end{eqnarray}
where
\begin{eqnarray}
l_{\star,\mu}(t)&=&{\cal A}_{\star,\mu}+\frac{({\cal A}_{\star,\mu+1}-{\cal A}_{\star,\mu})}{(\tau_{\mu+1}-\tau_\mu)}(t-\tau_\mu),\nonumber\\
\end{eqnarray}
and
\begin{eqnarray}
w(t,x,y)&=&\Theta(t-x)-\Theta(t-y).
\end{eqnarray}
Here, ${\cal A}_{\star,\mu}\equiv{\cal A}_{\star}(\tau_\mu)$.  This piecewise linear interpolation scheme has been used in \citet{hcm+12} to search for monopolar timing residual correlations associated with terrestrial clock errors and \citet{kcs+13} in an effort to account for time-variable dispersion measures in individual pulsar timing data sets. As we wish to incorporate this linear interpolant into the timing model, we will need to know how the timing residuals change as the parameters of the interpolant are changed, i.e., 
\begin{eqnarray}
\frac{\partial\delta t_K(t)}{\partial {\cal A}_{\star,\lambda}}&=&G_K^{\star}\left[1-\frac{t-\tau_1}{\tau_2-\tau_1}\right]w(t,\tau_1,\tau_2);\nonumber\\&&{\rm if}~\lambda=1,\nonumber\\\nonumber\\
&=&G_K^{\star}\left\{\frac{t-\tau_{\lambda-1}}{\tau_\lambda-\tau_{\lambda-1}}w(t,\tau_{\lambda-1},\tau_\lambda)+\right.\nonumber\\&&\left.\left[1-\frac{t-\tau_\lambda}{\tau_{\lambda+1}-\tau_\lambda}\right]w(t,\tau_\lambda,\tau_{\lambda+1})\right\};\nonumber\\&&{\rm if}~1<\lambda<N_\tau,\nonumber\\\nonumber\\
&=&G_K^{\star}\frac{t-\tau_{N_\tau-1}}{\tau_{N_\tau}-\tau_{N_\tau-1}}w(t,\tau_{N_\tau-1},\tau_{N_\tau});\nonumber\\&&{\rm if}~\lambda=N_\tau.
\end{eqnarray}

To estimate ${\cal A}_{+,\mu}$ and ${\cal A}_{\times,\mu}$, they must be fit for simultaneously with the timing models of several pulsars.  Without doing this, structure induced in the residuals of a single pulsar from a GW could be highly covariant with the structure anticipated from an inaccurate timing model.  The timing model parameters would become biased away from their maximum likelihood values and much of the power in the residuals from the GW would be absorbed.  To execute a simultaneous fit of all the pulsars in the PTA that can accommodate global parameters that are shared by all the timing models, we can construct modified design and noise covariance matrices and use standard least-squares fitting techniques as in Equations \ref{eq:5.6} and \ref{eq:5.7}.  A modified design matrix that allows us to carry out such a fit is ${\bf M_g}$, structured as follows:

\begin{eqnarray}
\left[\begin{array}{ccccc}
{\bf M_1}&\ldots&{\bf 0}&\frac{\partial\delta t_1({\bf t}_1)}{\partial{\cal A}_{+,1}}&\ldots~~~~~\\
\vdots&\ddots&\vdots&\vdots&\ddots~~~~~\\
{\bf 0}&\ldots&{\bf M_{N_P}}&\frac{\partial\delta t_{N_P}({\bf t}_{N_P})}{\partial{\cal A}_{+,1}}&\ldots~~~~~\\
\end{array}\right.\nonumber\\
\left.\begin{array}{cccc}
\frac{\partial\delta t_1({\bf t}_1)}{\partial{\cal A}_{+,N_\tau}}&\frac{\partial\delta t_1({\bf t}_1)}{\partial{\cal A}_{\times,1}}&\ldots&\frac{\partial\delta t_1({\bf t}_1)}{\partial{\cal A}_{\times,N_\tau}}\\
\vdots&\vdots&\ddots&\vdots\\
\frac{\partial\delta t_{N_P}({\bf t}_{N_P})}{\partial{\cal A}_{+,N_\tau}}&\frac{\partial\delta t_{N_P}({\bf t}_{N_P})}{\partial{\cal A}_{\times,1}}&\ldots&\frac{\partial\delta t_{N_P}({\bf t}_{N_P})}{\partial{\cal A}_{\times,N_\tau}}
\end{array}\right].\nonumber\\
\end{eqnarray}

If the $K^{\rm th}$ pulsar has a $n_K\times m_K$ design matrix, ${\bf M_g}$ will be a $(\sum_Kn_K)\times(2N_\tau+\sum_Km_K)$ matrix that is block diagonal except for the rightmost $2N_\tau$ columns. The maximum possible value of $N_\tau$ is set by the requirement that ${\bf M_g}$ have more rows than columns, or that there be more ToAs than model parameters; in practice we choose $N_\tau$ to be significantly smaller than this theoretical limit. The global noise covariance matrix will be $(\sum_Kn_K)\times(\sum_Kn_K)$ and block diagonal:
\begin{eqnarray}
{\bf C_g}=\left[\begin{array}{ccc}
{\bf C_1}&\ldots&{\bf 0}\\
\vdots&\ddots&\vdots\\
{\bf 0}&\ldots&{\bf C_{N_P}}
\end{array}\right].
\end{eqnarray}

The matrices ${\bf C_g}$ and ${\bf M_g}$ can be used along with a vector of all of the timing residuals, ${\bf \delta t}^T = [{\bf \delta t_1}^T\ldots{\bf\delta t_{N_P}}^T]$, in Equations \ref{eq:5.6} and \ref{eq:5.7} to carry out a global fit simultaneously for the timing models of all the pulsars in the array and the model for a quadrupolar GW signal coming from direction ${\bf \hat{n}}$.  However, without some additional conditioning, the fit will be ill-behaved because certain components of signals in $A_+$ and $A_\times$ will induce structure in each pulsar's timing residuals that will be indistinguishable from structure caused by incorrect estimates of that pulsar's timing model parameters, resulting in a singular parameter covariance matrix.

For example, suppose that $A_{+}(t)$ and $A_{\times}(t)$ are simply quadratics, $Q_+(t)$ and $Q_\times(t)$.  Then, according to Equation \ref{eq:5.3}, the timing perturbation from this GW in the $K^{\rm th}$ pulsar's residuals will be $\delta t_K^h(t)=G^+_KQ_+(t)+G^\times_KQ_\times(t)$, which is itself a simple quadratic.  The most fundamental parameters in any pulsar timing model are the spin parameters: a reference rotational phase, $\phi_{K}$, the pulsar's spin frequency, $f_K$, and the pulsar's spin-down rate, $\dot{f}_K$.  Errors in the timing model estimates of these parameters lead to quadratic structure in the timing residuals.  In fact, the only means we have of measuring these quantities for a particular pulsar is by fitting a quadratic to that pulsar's timing residuals.  If a GW were creating additional quadratic structure in a pulsar's timing residuals beyond what is expected from incorrect estimates of that pulsar's spin parameters, the two effects could never be meaningfully differentiated.  For this reason, if we are to simultaneously fit the timing models for all the pulsars in the PTA with models for $A_+$ and $A_\times$, we must constrain the models to not contain quadratics.  This can be implemented by enforcing these six equality constraints:
\begin{eqnarray}
\label{eq:5.14}
&&\sum_{\mu=1}^{N_\tau}{\cal A}_{\star,\mu}=0,\nonumber\\
&&\sum_{\mu=1}^{N_\tau}\tau_\mu{\cal A}_{\star,\mu}=0,\nonumber\\
&&\sum_{\mu=1}^{N_\tau}\tau_\mu^2{\cal A}_{\star,\mu}=0.
\end{eqnarray}
We can approximately satisfy these six constraints while simultaneously conducting the global fit by appending six appropriately constructed columns to ${\bf M_g}$, six zero-value pseudo-residuals to $\delta{\bf t}$, and six very small diagonal elements, ${\epsilon}$, to ${\bf C_g}$.  We find that this procedure is numerically stable when we make $\epsilon$ five to six orders of magnitude smaller than the largest diagonal element of ${\bf C_g}$.  Modern PTA data sets include pulsars with rms ToA uncertainty ranging between several tens of nanoseconds and a few microseconds.  Our condition on $\epsilon$ thus enforces that the linear equality constraints on ${\cal A}_+$ and ${\cal A}_\times$ are satisfied with a precision consistent with or in excess of the timing precisions achievable with the best-timed known pulsars.

In addition to the spin parameters of a pulsar, for the purposes of precision pulsar timing, the celestial coordinates of the pulsar are always fit for, and if the precision with which the pulsar can be timed warrants it, two proper motion terms and a parallax are also fit for.  The influence of these astrometric parameters on timing models principally concerns the calculation of the Roemer delay---the difference in light-travel-time for pulses arriving at Earth-based observatories and the solar-system barycenter.  Small estimation errors in the position parameters of a pulsar lead to annual sinusoidal fluctuations in the timing residuals of that pulsar.  The sinusoid can be of any phase.  Similarly, incorrect estimates for proper motion lead to annual sinusoidal fluctuations of any phase in a pulsar's residuals, but the amplitude of the sinusoid grows linearly in time as the expected pulsar position deviates more and more from the true pulsar position. 

Like a quadratic, if a GW were to induce sinusoidal structure with a one-year period (possibly with an amplitude changing linearly in time) in the residuals of any or all the pulsars in a PTA, that structure could not be differentiated from the signatures of incorrect estimates for the positions and proper motions of the affected pulsars.  We must thus also constrain ${\cal A}_+$ and ${\cal A}_\times$ to not contain such signatures:
\begin{eqnarray}
\label{eq:5.15}
&&\sum_{\mu=1}^{N_\tau}\sin{(\omega_1\tau_\mu)}{\cal A}_{\star,\mu}=0,\nonumber\\
&&\sum_{\mu=1}^{N_\tau}\cos{(\omega_1\tau_\mu)}{\cal A}_{\star,\mu}=0,\nonumber\\
&&\sum_{\mu=1}^{N_\tau}\tau_\mu\sin{(\omega_1\tau_\mu)}{\cal A}_{\star,\mu}=0,\nonumber\\
&&\sum_{\mu=1}^{N_\tau}\tau_\mu\cos{(\omega_1\tau_\mu)}{\cal A}_{\star,\mu}=0,
\end{eqnarray}
where $\omega_1=2\pi~{\rm yr}^{-1}$.

In the context of fitting linear interpolants to individual pulsar timing data sets to model time-variable dispersion measure, \citet{kcs+13} were the first to discuss and implement the constraints discussed above.  Because pulsar timing models often fit for parallax, errors in which induce sinusoidal oscillations with a half-year period in timing residuals, \citet{kcs+13} also constrained their linear interpolants to not contain such biannual sinusoids.  But, a parallax constraint is unnecessary for the global, multi-pulsar fits we are discussing.  Fitting for parallax in all the pulsars in an array does not generate problems when globally fitting for ${\cal A}_+$ and ${\cal A}_\times$ in the same way as fitting for the pulsars' rotational parameters, positions, or proper motions.  Estimation errors in the parallax induce sinusoidal structure in the timing residuals of a pulsar with a specific phase that depends only on the equatorial longitude of that pulsar.  If a GW were to induce biannual sinusoidal structure in a pulsar's residuals, it would have to have a very specific phase to potentially be confused as an error in the parallax estimate for that pulsar, but then, the residuals of other pulsars in the PTA would also show evidence of biannual sinusoidal fluctuations with the same phase in their timing residuals and with amplitudes that vary in accordance with the quadrupolar nature of a GW.  If they do, these biannual sinusoidal fluctuations can be reliably ascribed to the activity of a GW in the vicinity of Earth.  If they do not, the fluctuations can instead be ascribed to an estimation error in the parallax of one pulsar.  For these reasons, no additional constraints regarding parallax must be enforced on ${\cal A}_+$ and ${\cal A}_\times$. 

\section{Tempo2 Usage}

Here we provide details for how to make use of the routines described in this paper.   In almost all cases the ${\cal A}_+(t)$ and ${\cal A}_\times(t)$ time series will be included as part of a global fit.  The following parameters should therefore be provided in a global parameter file (which we assume is called ``apac.par"):

\begin{verbatim}
# The following line sets (S) the 
#quadrupolar (Q) interpolation function
#(IFUNC) for the plus (p) polarisation. 
#The Ò1Ó indicates the type of interpolation 
#(linear) and the Ò2Ó indicates that this 
#will be fitted globally to all pulsars
SQIFUNC_p 1 2
# The next line is the same for the 
#cross polarisation
SQIFUNC_c 1 2
# We now need to define the direction 
#(in radians) of the quadrupole for the 
#plus and cross functions 
#(usually they are the same)
QIFUNC_POS_P <ra> <dec>
QIFUNC_POS_c <ra> <dec>

# We now define the actual grid points 
#for the plus polarisation
QIFUNC_p1 <mjd> <val> <err>
QIFUNC_p2 <mjd> <val> <err>
É

# and also for the cross polarisation
QIFUNC_c1 <mjd> <val> <err>
QIFUNC_c2 <mjd> <val> <err>
É

# We then ensure that the constraints 
#are correctly applied
CONSTRAIN QIFUNC_p
CONSTRAIN QIFUNC_c
\end{verbatim}

Assuming that four pulsars have been observed, their parameters are in individual files (psr1.par, psr2.par, psr3.par and psr4.par) and their ToAs are in psr1.tim, psr2.tim etc., then the \textsc{tempo2} fit can be carried out using:

\begin{verbatim}
tempo2 -f psr1.par psr1.tim -f psr2.par
psr2.tim -f psr3.par psr3.tim -f psr4.par 
psr4.tim -global apac.par
\end{verbatim}

To include red noise models (defined in model.dat), use:

\begin{verbatim}
tempo2 -f psr1.par psr1.tim -f psr2.par 
psr2.tim -f psr3.par psr3.tim -f psr4.par 
psr4.tim -global apac.par -dcf model.dat
\end{verbatim}

The $\hat{\cal A}_+$ and $\hat{\cal A}_\times$ time series are written to files called ``aplus\_t2.dat" and ``across\_t2.dat". The covariance matrix ${\bf C}_{+\times}$ is written to a file called ``aplus\_across.cvm".


\bibliography{mzh+15}

\begin{thebibliography}{50}
\expandafter\ifx\csname natexlab\endcsname\relax\def\natexlab#1{#1}\fi

\bibitem[{{Arzoumanian} {et~al.}(2014){Arzoumanian}, {Brazier},
  {Burke-Spolaor}, {Chamberlin}, {Chatterjee}, {Cordes}, {Demorest}, {Deng},
  {Dolch}, {Ellis}, {Ferdman}, {Garver-Daniels}, {Jenet}, {Jones}, {Kaspi},
  {Koop}, {Lam}, {Lazio}, {Lommen}, {Lorimer}, {Luo}, {Lynch}, {Madison},
  {McLaughlin}, {McWilliams}, {Nice}, {Palliyaguru}, {Pennucci}, {Ransom},
  {Sesana}, {Siemens}, {Stairs}, {Stinebring}, {Stovall}, {Swiggum},
  {Vallisneri}, {van Haasteren}, {Wang}, {Zhu}, \& {NANOGrav
  Collaboration}}]{abb+14}
{Arzoumanian}, Z., {Brazier}, A., {Burke-Spolaor}, S., {et~al.} 2014, ApJ, 794,
  141

\bibitem[{{Arzoumanian} {et~al.}(2015{\natexlab{a}}){Arzoumanian}, {Brazier},
  {Burke-Spolaor}, {Chamberlin}, {Chatterjee}, {Christy}, {Cordes}, {Cornish},
  {Demorest}, {Deng}, {Dolch}, {Ellis}, {Ferdman}, {Fonseca}, {Garver-Daniels},
  {Jenet}, {Jones}, {Kaspi}, {Koop}, {Lam}, {Lazio}, {Levin}, {Lommen},
  {Lorimer}, {Luo}, {Lynch}, {Madison}, {McLaughlin}, {McWilliams}, {Nice},
  {Palliyaguru}, {Pennucci}, {Ransom}, {Siemens}, {Stairs}, {Stinebring},
  {Stovall}, {Swiggum}, {Vallisneri}, {van Haasteren}, {Wang}, {Zhu}, \&
  {NANOGrav Collaboration}}]{abb+15}
---. 2015{\natexlab{a}}, ApJ, 810, 150

\bibitem[{{Arzoumanian} {et~al.}(2015{\natexlab{b}}){Arzoumanian}, {Brazier},
  {Burke-Spolaor}, {Chamberlin}, {Chatterjee}, {Christy}, {Cordes}, {Cornish},
  {Demorest}, {Deng}, {Dolch}, {Ellis}, {Ferdman}, {Fonseca}, {Garver-Daniels},
  {Jenet}, {Jones}, {Kaspi}, {Koop}, {Lam}, {Lazio}, {Levin}, {Lommen},
  {Lorimer}, {Luo}, {Lynch}, {Madison}, {McLaughlin}, {McWilliams},
  {Mingarelli}, {Nice}, {Palliyaguru}, {Pennucci}, {Ransom}, {Sampson},
  {Sanidas}, {Sesana}, {Siemens}, {Simon}, {Stairs}, {Stinebring}, {Stovall},
  {Swiggum}, {Taylor}, {Vallisneri}, {van Haasteren}, {Wang}, \&
  {Zhu}}]{abb+15_2}
---. 2015{\natexlab{b}}, arXiv:1508.03024

\bibitem[{{Babak} {et~al.}(2015){Babak}, {Petiteau}, {Sesana}, {Brem},
  {Rosado}, {Taylor}, {Lassus}, {Hessels}, {Bassa}, {Burgay}, {Caballero},
  {Champion}, {Cognard}, {Desvignes}, {Gair}, {Guillemot}, {Janssen},
  {Karuppusamy}, {Kramer}, {Lazarus}, {Lee}, {Lentati}, {Liu}, {Mingarelli},
  {Oslowski}, {Perrodin}, {Possenti}, {Purver}, {Sanidas}, {Smits}, {Stappers},
  {Theureau}, {Tiburzi}, {van Haasteren}, {Vecchio}, \& {Verbiest}}]{bps+15}
{Babak}, S., {Petiteau}, A., {Sesana}, A., {et~al.} 2015, arXiv:1509.02165

\bibitem[{{Blandford} {et~al.}(1984){Blandford}, {Narayan}, \&
  {Romani}}]{brn84}
{Blandford}, R., {Narayan}, R., \& {Romani}, R.~W. 1984, Journal of
  Astrophysics and Astronomy, 5, 369

\bibitem[{{Champion} {et~al.}(2010){Champion}, {Hobbs}, {Manchester},
  {Edwards}, {Backer}, {Bailes}, {Bhat}, {Burke-Spolaor}, {Coles}, {Demorest},
  {Ferdman}, {Folkner}, {Hotan}, {Kramer}, {Lommen}, {Nice}, {Purver},
  {Sarkissian}, {Stairs}, {van Straten}, {Verbiest}, \& {Yardley}}]{chm+10}
{Champion}, D.~J., {Hobbs}, G.~B., {Manchester}, R.~N., {et~al.} 2010, ApJL,
  720, L201

\bibitem[{{Coles} {et~al.}(2011){Coles}, {Hobbs}, {Champion}, {Manchester}, \&
  {Verbiest}}]{chc+11}
{Coles}, W., {Hobbs}, G., {Champion}, D.~J., {Manchester}, R.~N., \&
  {Verbiest}, J.~P.~W. 2011, MNRAS, 418, 561

\bibitem[{{Cordes} \& {Jenet}(2012)}]{cj12}
{Cordes}, J.~M., \& {Jenet}, F.~A. 2012, ApJ, 752, 54

\bibitem[{{Cordes} \& {Shannon}(2010)}]{cs10}
{Cordes}, J.~M., \& {Shannon}, R.~M. 2010, arXiv:1010.3785

\bibitem[{{Cornish} \& {van Haasteren}(2014)}]{cv14}
{Cornish}, N.~J., \& {van Haasteren}, R. 2014, arXiv:1406.4511

\bibitem[{{Cutler} {et~al.}(2014){Cutler}, {Burke-Spolaor}, {Vallisneri},
  {Lazio}, \& {Majid}}]{cbv+14}
{Cutler}, C., {Burke-Spolaor}, S., {Vallisneri}, M., {Lazio}, J., \& {Majid},
  W. 2014, PhRvD, 89, 042003

\bibitem[{{Demorest} {et~al.}(2013){Demorest}, {Ferdman}, {Gonzalez}, {Nice},
  {Ransom}, {Stairs}, {Arzoumanian}, {Brazier}, {Burke-Spolaor}, {Chamberlin},
  {Cordes}, {Ellis}, {Finn}, {Freire}, {Giampanis}, {Jenet}, {Kaspi}, {Lazio},
  {Lommen}, {McLaughlin}, {Palliyaguru}, {Perrodin}, {Shannon}, {Siemens},
  {Stinebring}, {Swiggum}, \& {Zhu}}]{dfg+13}
{Demorest}, P.~B., {Ferdman}, R.~D., {Gonzalez}, M.~E., {et~al.} 2013, ApJ,
  762, 94

\bibitem[{{Detweiler}(1979)}]{d79}
{Detweiler}, S. 1979, ApJ, 234, 1100

\bibitem[{{Edwards} {et~al.}(2006){Edwards}, {Hobbs}, \& {Manchester}}]{ehm06}
{Edwards}, R.~T., {Hobbs}, G.~B., \& {Manchester}, R.~N. 2006, MNRAS, 372, 1549

\bibitem[{{Estabrook} \& {Wahlquist}(1975)}]{ew75}
{Estabrook}, F.~B., \& {Wahlquist}, H.~D. 1975, General Relativity and
  Gravitation, 6, 439

\bibitem[{{Folkner} {et~al.}(2008){Folkner}, {Williams}, \& {Boggs}}]{fwb08}
{Folkner}, W.~M., {Williams}, J.~G., \& {Boggs}, D.~H. 2008, JPL IOM
  343R-08-003

\bibitem[{{Foster} \& {Backer}(1990)}]{fb90}
{Foster}, R.~S., \& {Backer}, D.~C. 1990, ApJ, 361, 300

\bibitem[{{Gair} {et~al.}(2014){Gair}, {Romano}, {Taylor}, \&
  {Mingarelli}}]{grt+14}
{Gair}, J., {Romano}, J.~D., {Taylor}, S., \& {Mingarelli}, C.~M.~F. 2014,
  Physical Review D, 90, 082001

\bibitem[{{Gregory}(2010)}]{g10}
{Gregory}, P. 2010, {Bayesian Logical Data Analysis for the Physical Sciences
  (Cambridge: Cambridge University Press)}

\bibitem[{{Hellings} \& {Downs}(1983)}]{hd83}
{Hellings}, R.~W., \& {Downs}, G.~S. 1983, ApJL, 265, L39

\bibitem[{{Hobbs} {et~al.}(2010){Hobbs}, {Archibald}, {Arzoumanian}, {Backer},
  {Bailes}, {Bhat}, {Burgay}, {Burke-Spolaor}, {Champion}, {Cognard}, {Coles},
  {Cordes}, {Demorest}, {Desvignes}, {Ferdman}, {Finn}, {Freire}, {Gonzalez},
  {Hessels}, {Hotan}, {Janssen}, {Jenet}, {Jessner}, {Jordan}, {Kaspi},
  {Kramer}, {Kondratiev}, {Lazio}, {Lazaridis}, {Lee}, {Levin}, {Lommen},
  {Lorimer}, {Lynch}, {Lyne}, {Manchester}, {McLaughlin}, {Nice}, {Oslowski},
  {Pilia}, {Possenti}, {Purver}, {Ransom}, {Reynolds}, {Sanidas}, {Sarkissian},
  {Sesana}, {Shannon}, {Siemens}, {Stairs}, {Stappers}, {Stinebring},
  {Theureau}, {van Haasteren}, {van Straten}, {Verbiest}, {Yardley}, \&
  {You}}]{haa+10}
{Hobbs}, G., {Archibald}, A., {Arzoumanian}, Z., {et~al.} 2010, Classical and
  Quantum Gravity, 27, 084013

\bibitem[{{Hobbs} {et~al.}(2012){Hobbs}, {Coles}, {Manchester}, {Keith},
  {Shannon}, {Chen}, {Bailes}, {Bhat}, {Burke-Spolaor}, {Champion},
  {Chaudhary}, {Hotan}, {Khoo}, {Kocz}, {Levin}, {Oslowski}, {Preisig}, {Ravi},
  {Reynolds}, {Sarkissian}, {van Straten}, {Verbiest}, {Yardley}, \&
  {You}}]{hcm+12}
{Hobbs}, G., {Coles}, W., {Manchester}, R.~N., {et~al.} 2012, MNRAS, 427, 2780

\bibitem[{{Huerta} {et~al.}(2015){Huerta}, {McWilliams}, {Gair}, \&
  {Taylor}}]{hmg+15}
{Huerta}, E.~A., {McWilliams}, S.~T., {Gair}, J.~R., \& {Taylor}, S.~R. 2015,
  arXiv:1504.00928

\bibitem[{{Keith} {et~al.}(2013){Keith}, {Coles}, {Shannon}, {Hobbs},
  {Manchester}, {Bailes}, {Bhat}, {Burke-Spolaor}, {Champion}, {Chaudhary},
  {Hotan}, {Khoo}, {Kocz}, {Os{\l}owski}, {Ravi}, {Reynolds}, {Sarkissian},
  {van Straten}, \& {Yardley}}]{kcs+13}
{Keith}, M.~J., {Coles}, W., {Shannon}, R.~M., {et~al.} 2013, MNRAS, 429, 2161

\bibitem[{{Kramer} \& {Champion}(2013)}]{kc13}
{Kramer}, M., \& {Champion}, D.~J. 2013, Classical and Quantum Gravity, 30,
  224009

\bibitem[{{Lee}(2013)}]{l13}
{Lee}, K.~J. 2013, Classical and Quantum Gravity, 30, 224016

\bibitem[{{Lee} {et~al.}(2011){Lee}, {Wex}, {Kramer}, {Stappers}, {Bassa},
  {Janssen}, {Karuppusamy}, \& {Smits}}]{lwk+11}
{Lee}, K.~J., {Wex}, N., {Kramer}, M., {et~al.} 2011, MNRAS, 414, 3251

\bibitem[{{Lentati} {et~al.}(2015){Lentati}, {Taylor}, {Mingarelli}, {Sesana},
  {Sanidas}, {Vecchio}, {Caballero}, {Lee}, {van Haasteren}, {Babak}, {Bassa},
  {Brem}, {Burgay}, {Champion}, {Cognard}, {Desvignes}, {Gair}, {Guillemot},
  {Hessels}, {Janssen}, {Karuppusamy}, {Kramer}, {Lassus}, {Lazarus}, {Liu},
  {Os{\l}owski}, {Perrodin}, {Petiteau}, {Possenti}, {Purver}, {Rosado},
  {Smits}, {Stappers}, {Theureau}, {Tiburzi}, \& {Verbiest}}]{ltm+15}
{Lentati}, L., {Taylor}, S.~R., {Mingarelli}, C.~M.~F., {et~al.} 2015,
  arXiv:1504.03692

\bibitem[{{Madison} {et~al.}(2013){Madison}, {Chatterjee}, \& {Cordes}}]{mcc13}
{Madison}, D.~R., {Chatterjee}, S., \& {Cordes}, J.~M. 2013, ApJ, 777, 104

\bibitem[{{Madison} {et~al.}(2014){Madison}, {Cordes}, \& {Chatterjee}}]{mcc14}
{Madison}, D.~R., {Cordes}, J.~M., \& {Chatterjee}, S. 2014, ApJ, 788, 141

\bibitem[{{Manchester}(2013)}]{m13_1}
{Manchester}, R.~N. 2013, Classical and Quantum Gravity, 30, 224010

\bibitem[{{Manchester} {et~al.}(2013){Manchester}, {Hobbs}, {Bailes}, {Coles},
  {van Straten}, {Keith}, {Shannon}, {Bhat}, {Brown}, {Burke-Spolaor},
  {Champion}, {Chaudhary}, {Edwards}, {Hampson}, {Hotan}, {Jameson}, {Jenet},
  {Kesteven}, {Khoo}, {Kocz}, {Maciesiak}, {Oslowski}, {Ravi}, {Reynolds},
  {Sarkissian}, {Verbiest}, {Wen}, {Wilson}, {Yardley}, {Yan}, \&
  {You}}]{mhb+13}
{Manchester}, R.~N., {Hobbs}, G., {Bailes}, M., {et~al.} 2013, PASA, 30, 17

\bibitem[{{McLaughlin}(2013)}]{m13_2}
{McLaughlin}, M.~A. 2013, Classical and Quantum Gravity, 30, 224008

\bibitem[{{Milosavljevi{\'c}} \& {Merritt}(2003)}]{mm03}
{Milosavljevi{\'c}}, M., \& {Merritt}, D. 2003, ApJ, 596, 860

\bibitem[{{Pshirkov} {et~al.}(2010){Pshirkov}, {Baskaran}, \&
  {Postnov}}]{pbp10}
{Pshirkov}, M.~S., {Baskaran}, D., \& {Postnov}, K.~A. 2010, MNRAS, 402, 417

\bibitem[{{Ravi} {et~al.}(2014){Ravi}, {Wyithe}, {Shannon}, {Hobbs}, \&
  {Manchester}}]{rws+14}
{Ravi}, V., {Wyithe}, J.~S.~B., {Shannon}, R.~M., {Hobbs}, G., \& {Manchester},
  R.~N. 2014, MNRAS, 442, 56

\bibitem[{{Sesana} \& {Vecchio}(2010)}]{sv10}
{Sesana}, A., \& {Vecchio}, A. 2010, Classical and Quantum Gravity, 27, 084016

\bibitem[{{Shannon} \& {Cordes}(2010)}]{sc10}
{Shannon}, R.~M., \& {Cordes}, J.~M. 2010, ApJ, 725, 1607

\bibitem[{{Shannon} {et~al.}(2013){Shannon}, {Ravi}, {Coles}, {Hobbs}, {Keith},
  {Manchester}, {Wyithe}, {Bailes}, {Bhat}, {Burke-Spolaor}, {Khoo}, {Levin},
  {Oslowski}, {Sarkissian}, {van Straten}, {Verbiest}, \& {Want}}]{src+13}
{Shannon}, R.~M., {Ravi}, V., {Coles}, W.~A., {et~al.} 2013, Science, 342, 334

\bibitem[{{Shannon} {et~al.}(2015){Shannon}, {Ravi}, {Lentati}, {Lasky},
  {Hobbs}, {Kerr}, {Manchester}, {Coles}, {Levin}, {Bailes}, {Bhat},
  {Burke-Spolaor}, {Dai}, {Keith}, {Os{\l}owski}, {Reardon}, {van Straten},
  {Toomey}, {Wang}, {Wen}, {Wyithe}, \& {Zhu}}]{srl+15}
{Shannon}, R.~M., {Ravi}, V., {Lentati}, L.~T., {et~al.} 2015, arXiv:1509.07320

\bibitem[{{Simon} {et~al.}(2014){Simon}, {Polin}, {Lommen}, {Stappers}, {Finn},
  {Jenet}, \& {Christy}}]{spl+14}
{Simon}, J., {Polin}, A., {Lommen}, A., {et~al.} 2014, ApJ, 784, 60

\bibitem[{{Standish}(2006)}]{s06}
{Standish}, Jr., E.~M. 2006, JPL IOM 343R-06-002

\bibitem[{{Taylor} {et~al.}(2015){Taylor}, {Mingarelli}, {Gair}, {Sesana},
  {Theureau}, {Babak}, {Bassa}, {Brem}, {Burgay}, {Caballero}, {Champion},
  {Cognard}, {Desvignes}, {Guillemot}, {Hessels}, {Janssen}, {Karuppusamy},
  {Kramer}, {Lassus}, {Lazarus}, {Lentati}, {Liu}, {Os{\l}owski}, {Perrodin},
  {Petiteau}, {Possenti}, {Purver}, {Rosado}, {Sanidas}, {Smits}, {Stappers},
  {Tiburzi}, {van Haasteren}, {Vecchio}, {Verbiest}, \& {EPTA
  Collaboration}}]{tmg+15}
{Taylor}, S.~R., {Mingarelli}, C.~M.~F., {Gair}, J.~R., {et~al.} 2015, Physical
  Review Letters, 115, 041101

\bibitem[{{Tiburzi} {et~al.}(2015){Tiburzi}, {Hobbs}, {Kerr}, {Coles}, {Dai},
  {Manchester}, {Possenti}, {Shannon}, \& {You}}]{thk+15}
{Tiburzi}, C., {Hobbs}, G., {Kerr}, M., {et~al.} 2015, arXiv:1510.02363

\bibitem[{{van Haasteren} \& {Levin}(2010)}]{vl10}
{van Haasteren}, R., \& {Levin}, Y. 2010, MNRAS, 401, 2372

\bibitem[{{van Haasteren} {et~al.}(2011){van Haasteren}, {Levin}, {Janssen},
  {Lazaridis}, {Kramer}, {Stappers}, {Desvignes}, {Purver}, {Lyne}, {Ferdman},
  {Jessner}, {Cognard}, {Theureau}, {D'Amico}, {Possenti}, {Burgay},
  {Corongiu}, {Hessels}, {Smits}, \& {Verbiest}}]{vlj+11}
{van Haasteren}, R., {Levin}, Y., {Janssen}, G.~H., {et~al.} 2011, MNRAS, 414,
  3117

\bibitem[{{Vigeland} \& {Vallisneri}(2014)}]{vv14}
{Vigeland}, S.~J., \& {Vallisneri}, M. 2014, MNRAS, 440, 1446

\bibitem[{{Wang} {et~al.}(2015){Wang}, {Hobbs}, {Coles}, {Shannon}, {Zhu},
  {Madison}, {Kerr}, {Ravi}, {Keith}, {Manchester}, {Levin}, {Bailes}, {Bhat},
  {Burke-Spolaor}, {Dai}, {Os{\l}owski}, {van Straten}, {Toomey}, {Wang}, \&
  {Wen}}]{whc+15}
{Wang}, J.~B., {Hobbs}, G., {Coles}, W., {et~al.} 2015, MNRAS, 446, 1657

\bibitem[{{Wang} {et~al.}(2014){Wang}, {Mohanty}, \& {Jenet}}]{wmj14}
{Wang}, Y., {Mohanty}, S.~D., \& {Jenet}, F.~A. 2014, ApJ, 795, 96

\bibitem[{{Zhu} {et~al.}(2014){Zhu}, {Hobbs}, {Wen}, {Coles}, {Wang},
  {Shannon}, {Manchester}, {Bailes}, {Bhat}, {Burke-Spolaor}, {Dai}, {Keith},
  {Kerr}, {Levin}, {Madison}, {Os{\l}owski}, {Ravi}, {Toomey}, \& {van
  Straten}}]{zhw+14}
{Zhu}, X.-J., {Hobbs}, G., {Wen}, L., {et~al.} 2014, MNRAS, 444, 3709

\end{thebibliography}
\end{document}